# Discovering High-Entropy Oxides with a Machine-Learning Interatomic Potential

*Jacob T. Sivak[1], Saeed S. I. Almishal[2], Mary K. Caucci[1], Yueze Tan[2], Dhiya Srikanth[2], Matthew Furst[2], Long-Qing Chen[2,3,4], Christina M. Rost[5], Jon-Paul Maria[2], *Susan B. Sinnott[1,2,6]

[1]*Department of Chemistry, The Pennsylvania State University, University Park, PA 16802, USA*
[2]*Department of Materials Science and Engineering, The Pennsylvania State University, University Park, PA 16802, USA*
[3]*Department of Mathematics, The Pennsylvania State University, University Park, PA 16802, USA*
[4]*Department of Engineering Science and Mechanics, The Pennsylvania State University, University Park, PA 16802, USA*
[5]*Department of Materials Science and Engineering, Virginia Polytechnic Institute and State University, Blacksburg, VA 24061, USA*
[6]*Institute for Computational and Data Sciences, The Pennsylvania State University, University Park, PA 16802, USA*

**Corresponding Authors:** Jacob T. Sivak jts6114@psu.edu and Susan B. Sinnott sbs5563@psu.edu

**Keywords:** high-entropy oxides, disorder, machine-learning interatomic potential, density functional theory, bulk synthesis

## Abstract

High-entropy materials shift the traditional materials discovery paradigm to one that leverages disorder, enabling access to unique chemistries unreachable through enthalpy alone. We present a self-consistent approach integrating computation and experiment to understand and explore single-phase rock salt high-entropy oxides. By leveraging a machine-learning interatomic potential, we rapidly and accurately map high-entropy composition space using our two descriptors: bond length distribution and mixing enthalpy. The single-phase stabilities for all experimentally stabilized rock salt compositions are correctly resolved, with dozens more compositions awaiting discovery.

Exploiting disorder in multicomponent materials enables access to an expansive, largely unexplored discovery space that is unattainable through enthalpic stabilization alone. Stabilizing materials through entropy pushes the solid solution paradigm to the extreme by incorporating multiple elements at random on a single sublattice. In turn, this boosts the configurational entropy making it possible at sufficiently high temperatures to form a single-phase multicomponent material rather than the enthalpically-preferred constituent components or an ensemble of ternary intermediates. This approach was originally shown in 2004 when Yeh and Cantor independently demonstrated the formation of high-entropy alloys containing five or six principal components [1,2]. Rost et al. then extended this methodology to ceramic systems in 2015 by illustrating the entropic stabilization of the $Mg_{1/5}Co_{1/5}Ni_{1/5}Cu_{1/5}Zn_{1/5}O$ rock salt high-entropy oxide (HEO) [3]. HEOs immediately stimulated interest from the materials community as promising characteristics were shown over a broad application spectrum, particularly enhanced cycling stability for energy storage [4]. The search for HEOs in various other crystal structures such as perovskite [5,6] and fluorite [7,8] has ensued, motivated by emerging functional property opportunities. However, the intrinsic disorder of HEOs results in a combinatorically difficult material space to traverse, especially if off-stoichiometric cation proportions are considered. Understanding how individual components act to either stabilize or destabilize the desired high-entropy phase, and their effect on resulting properties, is essential to advance the discovery and accelerate the implementation of these highly disordered materials.

Designing high-entropy materials *in-silico* presents a formidable challenge as the inherent disorder requires navigating through a complex space, making high-throughput density functional theory (DFT) predictions daunting, even with current computing capabilities. Formulating descriptors that capture single-phase many-component formulation spaces are also necessary as enthalpy is no longer the predominant thermodynamic driver; in such cases, formalisms like convex hulls alone may prove insufficient. The entropy-forming ability (EFA) descriptor has been proposed for high-entropy carbide discovery, relating the energy distribution to single-phase stability: a broader distribution is more energetically expensive to introduce configurational disorder into the system leading to a lower EFA [9]. A recent extension to EFA was provided through an additional enthalpic contribution to form the so-called disordered enthalpy-entropy descriptor (DEED) [10], and was used to screen novel high-entropy boride and carbonitride compositions. While successful across different ceramic materials, this framework requires many DFT calculations. Descriptors for HEOs specifically are limited to those which approximate HEOs as averages of their $A_{1/2}A'_{1/2}O_x$ constituents [11,12]. The average and standard deviation of these approximated mixing enthalpies rationalize the single-phase formation of the prototypical rock salt HEO, $Mg_{1/5}Co_{1/5}Ni_{1/5}Cu_{1/5}Zn_{1/5}O$, as well the recently synthesized Mn and Fe containing $Mg_{1/5}Mn_{1/5}Fe_{1/5}Co_{1/5}Ni_{1/5}O$ [13]. While the two-cation constituents offer an interpretable picture of HEO stability that is computationally modest, the unique characteristics of high-component HEOs may not be captured within this simplified framework.

To build on this foundation, this manuscript explores the following hypothesis: the single-phase stability of a many-cation rock salt oxide crystal with arbitrary cation ratios can be predicted by combining (1) the distribution of relaxed first near-neighbor bond lengths, and (2) the mixing enthalpy relative to the $A^{2+}O^{2-}$ ground states. These quantities are descriptors which can be calculated for thousands of compositions within the parent formulation envelope with an extremely low computational expense by leveraging recent advances in machine-learning interatomic potentials. Plotting these descriptors on cartesian coordinates will create a phase diagram, whose boundaries between single- and multi-phase zones are identified by experimentally synthesizing a subset of formulations to identify compositional solid solution thresholds. The following sections detail the combined computational and experimental framework to test this proposal.

We begin this process by defining the mixing enthalpy as the energy of a given HEO ($E[HEO]$) relative to all constituent $A^{2+}O^{2-}$ ground states ($E[A_i^{2+}O^{2-}]$):

$$\Delta H_{mix} = E[HEO] - \sum_i n_i E[A_i^{2+}O^{2-}] \tag{1}$$

where $n_i$ is the relative fraction of each end member in the HEO. As a result of only considering $A^{2+}O^{2-}$ competing phases, $\Delta H_{mix}$ is most valid for experimental conditions in which cations are stable at 2+ valence. As rock salt HEOs have been shown to be sensitive to the chemical potential of oxygen during synthesis [13,14], we choose this quantity rather than the enthalpy above hull or decomposition enthalpy [15]. These quantities may ultimately be more appropriate to describe non-equilibrium synthesis of HEOs [14]. $\Delta H_{mix}$ can be considered as the enthalpic barrier to rock salt HEO formation, hence a smaller value should promote single-phase rock salt formation. Our second descriptor, the bond length distribution, is quantified using the standard deviation of relaxed first near-neighbor (NN) cation-anion bond lengths:

$$\sigma_{bonds} = \sqrt{\frac{\sum_i (a_i - \bar{a})^2}{N}} \tag{2}$$

where $a_i$ is each bond length, $\bar{a}$ is the average bond lengths, and $N$ is the total number of bond lengths. A narrower distribution of bond lengths, and therefore a lower value for $\sigma_{bonds}$, indicates minimal lattice distortion, hence should more easily form a single-phase rock salt HEO. $\sigma_{bonds}$ is inherently similar to the Hume-Rothery rule that differences in atomic sizes govern solid solubility of alloys [16]. Additionally, a correlation has recently been identified between the local lattice distortion and EFA descriptor used for the computational discovery of high-entropy carbides [17].

While DFT is capable of calculating these two descriptors, the intrinsic disorder of HEOs and other high-entropy materials necessitates a significant computational investment. Most studies are inhibited to explore only the fully disordered limit with a single special quasi-random structure (SQS) of a large supercell [18], or a multitude of smaller symmetrically unique cells [19]. In both cases, the cubic scaling of DFT results in a heavy computational burden for even a single high-entropy composition. To overcome this computational cost, we employ a machine-learning interatomic potential (MLIP), specifically the Crystal Hamiltonian Graph Neural Network (CHGNet) [20], which provides a universal potential nearing the accuracy of DFT [21]. By fitting magnetic moments, energies, forces, and stresses in a crystal graph convolutional neural network, this MLIP approaches near-DFT accuracy with mean absolute errors (MAEs) of only 30 meV/atom and 77 meV/Å for energy and force, respectively, making it one of the most accurate universal MLIPs to date [20,21]. Providing both a structural and charge representation within its construction, CHGNet provides an ideal engine for high-throughput screening of HEOs and other high-entropy materials that are prohibitively expensive to study solely with DFT.

We first benchmark the pretrained, universal CHGNet MLIP against DFT calculations to investigate the accuracy for rock salt HEOs; multicomponent HEOs are not fitted explicitly as the training data largely consists of ordered systems. We construct our benchmarking data set from the cation cohort composed of Mg, Ca, Mn, Fe, Co, Ni, Cu, and Zn to form 160-atom rock salt SQSs of the 56 ($_8C_5$) possible five-cation equimolar combinations. Parameters consistent with the Materials Project [22,23] are utilized in DFT calculations to ensure compatibility with the calculations on which CHGNet was trained. We focus here only on the two quantities necessary for our proposed descriptors: the total energy, and the relaxed first NN bond lengths between metal and oxygen ions. Figure 1a shows a parity plot between DFT and CHGNet predicted total energies, resulting in a MAE of only 18.7 meV/atom. At nearly 38% less than the MAE in the original report by Deng et al. for materials spanning the entire periodic table [20], this result suggests that the high-entropy rock salt oxides investigated here may be more accurately described by CHGNet than other compositions in the original testing data set. The relaxed first NN bond lengths predicted by CHGNet compared to DFT for all HEO compositions explored are shown in Figure 1b. CHGNet accurately predicts the local structure for our rock salt HEOs as evident by the relatively high degree of linearity ($R^2$ = 0.64) as well a low MAE of 0.041 Å for all 26,880 bond lengths investigated. Importantly, the error distribution with respect to DFT is found to be normal and centered around zero. We further investigate our bond length benchmarking set by comparing compositions with and without Cu, as the $CuO_6$ octahedra exhibits local tetragonal distortions via a Jahn-Teller (JT) mechanism [24,25]. Larger errors compared to DFT are found for HEOs containing Cu (Figure S1a, Supplemental Material [26]), while minimal errors are observed for HEOs without Cu (Figure S1b, Supplemental Material [26]) resulting in a reduced MAE of only 0.018 Å and an increased $R^2$ value of 0.90. As the JT distortion of the $CuO_6$ octahedra are indeed captured by the CHGNet MLIP (Figure S2, Supplemental Material [26]), we attribute the larger error for Cu-containing compositions as a result of the anisotropic JT distortions rather than the inability of CHGNet to capture these electronically-driven local distortions. We also note that CHGNet is capable of predicting the parent oxide ground state structures with high accuracy (Table S1).

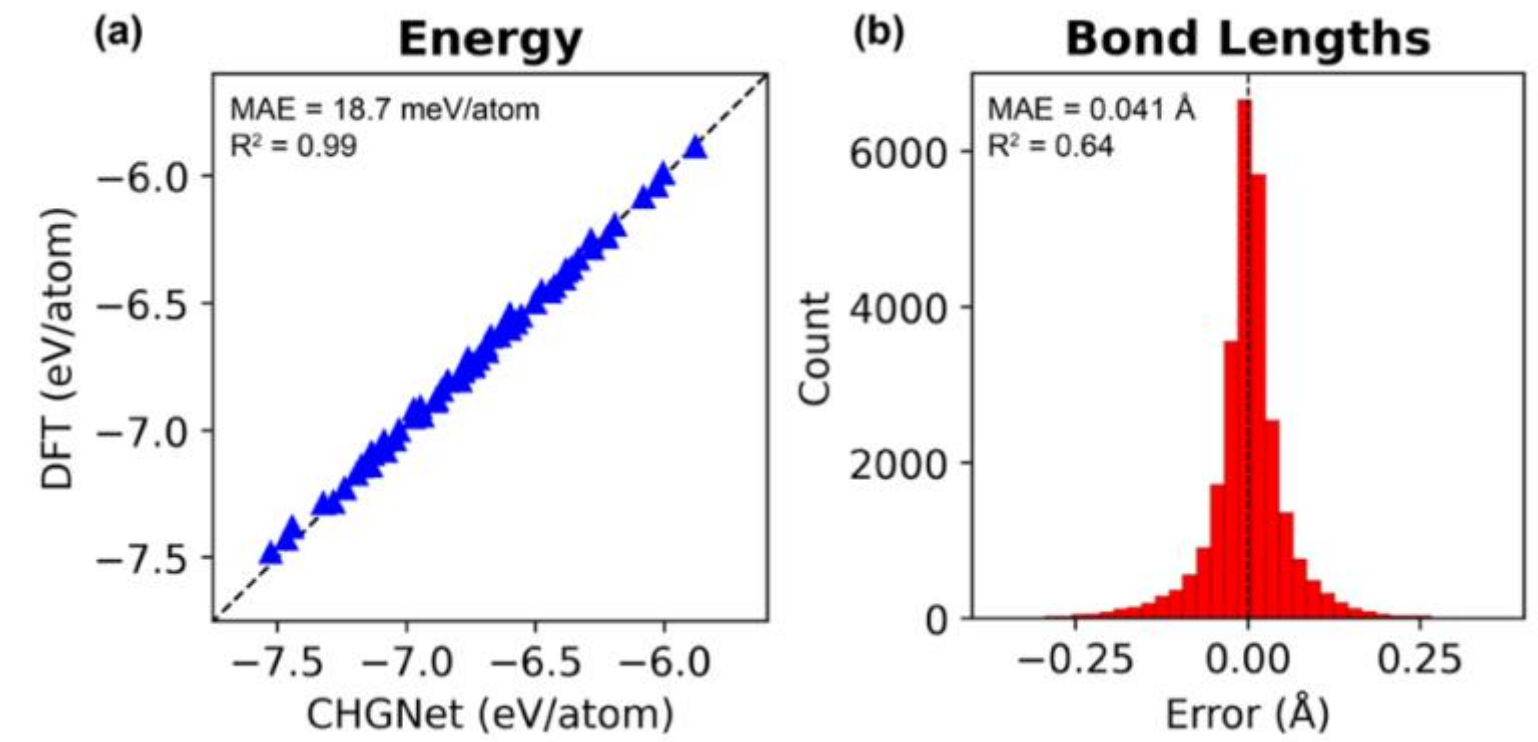

Figure 1. Benchmarking data for CHGNet compared to DFT for (a) total energy and (b) relaxed first-near neighbor bond lengths of equimolar five-cation rock salt oxides. The x-axis for (b) is the difference between DFT and CHGNet bond lengths.

Having established the accuracy of CHGNet, we leverage this MLIP as our computational engine to calculate our two descriptors, $\Delta H_{mix}$ and $\sigma_{bonds}$, for the $A_{1/2}A'_{1/2}O$ compositions of our cation cohort. Exploring trends for the two-cation rock salts allows a fundamental understanding of how different cation combinations affect descriptor values. For adequate sampling, 10 randomly decorated 1200-atom supercells are used for each $A_{1/2}A'_{1/2}O$ composition. Heatmaps of our descriptors are shown in Figure 2 in which darker shades indicate larger values for both descriptors and therefore represent cation combinations predicted to be more difficult to form a single-phase rock salt HEO. The largest value for $\Delta H_{mix}$ is observed for $Cu_{1/2}Zn_{1/2}O$ with an enthalpic penalty of 149 meV/atom, reflecting negligible rock salt solubility of CuO tenorite and ZnO wurtzite as seen from the phase diagrams [3,14]. Larger values for many of the other Cu and Zn containing cation combinations are observed as well. Conversely, $Mg_{1/2}Ni_{1/2}O$ has the smallest $\Delta H_{mix}$ value (34 meV/atom), agreeing with their ability to form a complete solid solution [14]. Ca combinations also result in larger $\Delta H_{mix}$ values, which we attribute to the ~30% increased size compared to the other cations [16] and is also reflected in increased local distortions captured by $\sigma_{bonds}$ (Figure 2b). Cu-containing combinations also exhibit larger distortions due to the aforementioned JT distortion. As a result, $Ca_{1/2}Cu_{1/2}O$ has the largest value for $\sigma_{bonds}$ of 0.31 Å. A larger $\sigma_{bonds}$ value for $Cu_{1/2}Zn_{1/2}O$ is observed due to relaxation away from the rock salt symmetry, which we attribute to the rock salt phase being above the convex hull for both CuO and ZnO. Minimal lattice distortion is observed for the other cation combinations.

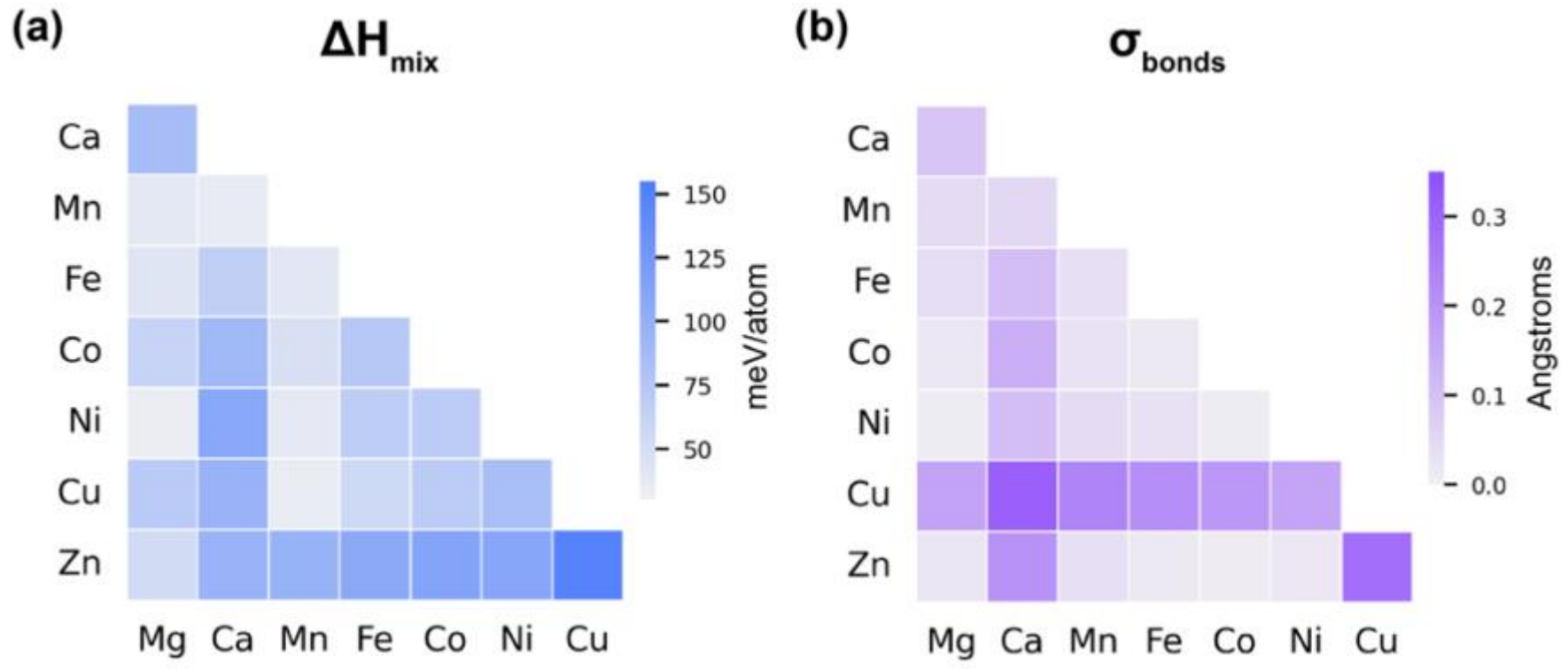

Figure 2. Heat maps of constituent $A_{1/2}A'_{1/2}O$ compositions for explored HEO compositions quantified using (a) $\Delta H_{mix}$ and (b) $\sigma_{bonds}$.

Next, we map high-entropy composition space for all equimolar 4-, 5- and 6-cation rock salt combinations using our descriptors. 154 equimolar HEO combinations are explored in total. The reduced computational cost of CHGNet enables a much more comprehensive sampling of atomic configurations than is attainable with DFT calculations: 50 randomly decorated 1200-atom supercells are used for each composition, resulting in 180,000 bond lengths per composition. A multitude of randomly decorated supercells are leveraged to more broadly sample configuration space rather than a single SQS as SQSs are designed to explore the high-temperature regime with full cation disorder; the structures in our data set have no indications of short-range ordering (Figure S3). We combine our descriptors with bulk ceramic synthesis and characterization to define single-phase stability thresholds for rock salt HEOs using eight compositions: the prototypical HEO $Mg_{1/5}Co_{1/5}Ni_{1/5}Cu_{1/5}Zn_{1/5}O$ [3], the five four-cation derivatives of $Mg_{1/5}Co_{1/5}Ni_{1/5}Cu_{1/5}Zn_{1/5}O$, $Mg_{1/6}Ca_{1/6}Co_{1/6}Ni_{1/6}Cu_{1/6}Zn_{1/6}O$, and $Mg_{1/5}Mn_{1/5}Fe_{1/5}Co_{1/5}Ni_{1/5}O$ [13]. The result of this process is provided in Figure 3 in which four HEOs are found to be single-phase (green circles), and four as multiple phases (red crosses). X-ray diffraction (XRD) patterns for the five four-cation derivatives of $Mg_{1/5}Co_{1/5}Ni_{1/5}Cu_{1/5}Zn_{1/5}O$ are shown in Figure S4 (see Supplemental Material [26]). Single-phase stability thresholds are defined using the largest descriptor values for experimentally realized single-phase compositions: $\Delta H_{mix}$ = 90.6 meV/atom for $Mg_{1/5}Co_{1/5}Ni_{1/5}Cu_{1/5}Zn_{1/5}O$ and $\sigma_{bonds}$ = 0.099 Å for $Mg_{1/4}Co_{1/4}Ni_{1/4}Cu_{1/4}O$. Importantly, CHGNet MLIP calculated $\Delta H_{mix}$ and $\sigma_{bonds}$ are able to correctly distinguish between single- and multi-phase compositions for all experimentally investigated equimolar rock salt HEOs of this cation cohort. Comparing the four single-phase HEOs, a reduced value for $\sigma_{bonds}$ is observed for $Mg_{1/4}Co_{1/4}Ni_{1/4}Zn_{1/4}O$ and $Mg_{1/5}Mn_{1/5}Fe_{1/5}Co_{1/5}Ni_{1/5}O$ compared to the prototypical HEO $Mg_{1/5}Co_{1/5}Ni_{1/5}Cu_{1/5}Zn_{1/5}O$ as Cu is no longer present. Similarly, $Mg_{1/4}Co_{1/4}Ni_{1/4}Cu_{1/4}O$ contains a larger fraction of Cu than its parent $Mg_{1/5}Co_{1/5}Ni_{1/5}Cu_{1/5}Zn_{1/5}O$, resulting in a larger value for $\sigma_{bonds}$. These reflect the intuition gained by the $A_{1/2}A'_{1/2}O$ descriptors in Figure 2 that the JT distortion for Cu-containing compositions increases the magnitude of local distortions. We also compare the CHGNet predicted local distortions to the Shannon ionic radii for all 154 compositions explored (Figure S5, Supplemental Material [26]) [27]. For HEO compositions without Cu, $\sigma_{bonds}$ and the standard deviation of ionic radii are correlated to one another, while for Cu-containing compositions the ionic radii significantly underestimates local lattice distortions as the JT distorted $CuO_6$ octahedra is not captured. This emphasizes the need for computational tools in HEO screening like CHGNet that capture electronically-driven effects such as a JT distortion or charge compensation. We also note that $Mg_{1/5}Mn_{1/5}Fe_{1/5}Co_{1/5}Ni_{1/5}O$ is predicted as single-phase in our approach despite being prepared using chloride precursors by Pu et al. as $\Delta H_{mix}$ is valid for conditions in which $Mn^{2+}$ and $Fe^{2+}$ are the most stable oxidation state [13]. A significant number of unexplored equimolar HEO compositions are predicted to form single-phase rock salt HEOs (Table S1, Supplemental Material [26]), many of which contain Mn and/or Fe. As a result, a multitude of rock salt HEOs await discovery once synthesis conditions are identified that inhibit the formation of higher oxidation states.

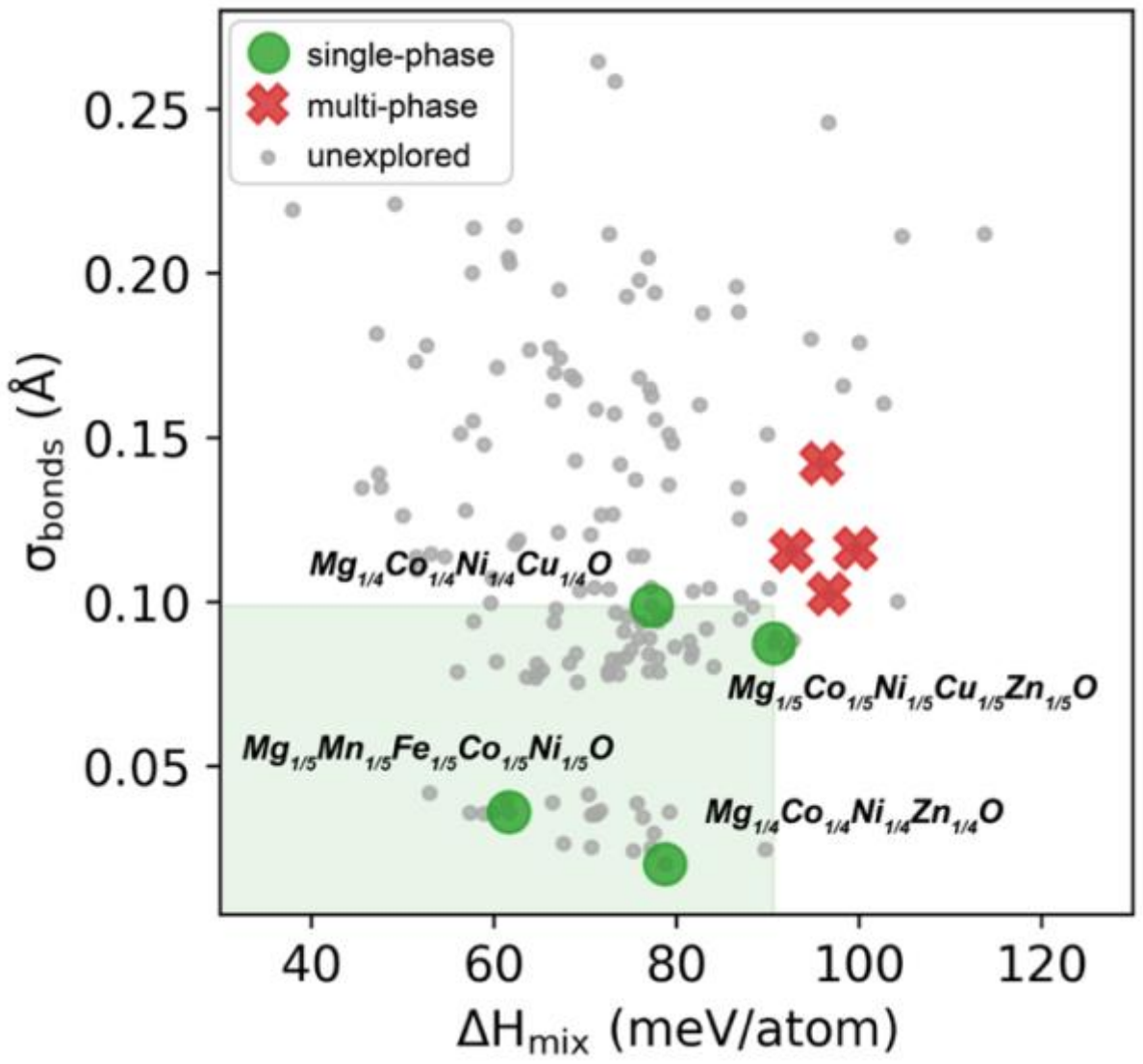

Figure 3. Map of rock salt HEO composition space using $\Delta H_{mix}$ and $\sigma_{bonds}$. Experimental synthesis results for single- and multi-phase are indicated with green circle and red crosses, respectively. Region predicted as single-phase shown in a green shade.

Inspired by the success of our approach for equimolar rock salt HEOs, we expand our discovery space by considering formulations with non-equimolar cation proportions. While equimolar stoichiometries are commonly explored to maximize configurational entropy, non-equimolar HEOs offer the benefit of additional chemical tunability, and therefore also property tunability. Specifically, we explore the composition space between the single-phase $Mg_{1/5}Co_{1/5}Ni_{1/5}Cu_{1/5}Zn_{1/5}O$ and multi-phase $Mg_{1/6}Ca_{1/6}Co_{1/6}Ni_{1/6}Cu_{1/6}Zn_{1/6}O$. Starting with the equimolar 6-cation $Mg_{1/6}Ca_{1/6}Co_{1/6}Ni_{1/6}Cu_{1/6}Zn_{1/6}O$, we map the effect of decreasing Ca concentration with $\Delta H_{mix}$ and $\sigma_{bonds}$, keeping the remaining cations in equimolar proportions. The smooth transition between these HEO compositions captured by CHGNet is depicted with blue circles in Figure 4a. It is important to note that none of these non-equimolar compositions cross within the single-phase stability region defined by the equimolar HEO compositions (Figure 3). We therefore carry out bulk synthesis for a series of non-equimolar HEOs at ambient conditions in this composition space and characterize the resulting crystal structure, number of phases, and cation proportions. The powder diffraction patterns in Figure 4b indicate the rock salt CaO peaks persist with decreasing Ca content down to 6.2% before completely dissolving into the high-entropy matrix at 4.0% Ca. We confirm all cation concentrations with X-ray fluorescence spectroscopy and monitor mass loss before and after every state of ceramic preparation. The composition for the single-phase HEO is then passed to our computational descriptors to self-consistently refine the single-phase stability thresholds, as noted by dashed black lines in Figure 4a ($\Delta H_{mix}$ = 92.2 meV/atom, $\sigma_{bonds}$ = 0.102 Å). Additional equimolar compositions are predicted to form a single-phase HEO as a result of this threshold refinement. Collectively, the reduced computational cost offered by CHGNet enables our collaborative screening architecture to logically explore high-entropy compositions that deviate from equimolarity, enabling access to a composition space in which properties can be finely tuned by altering relative cation proportions.

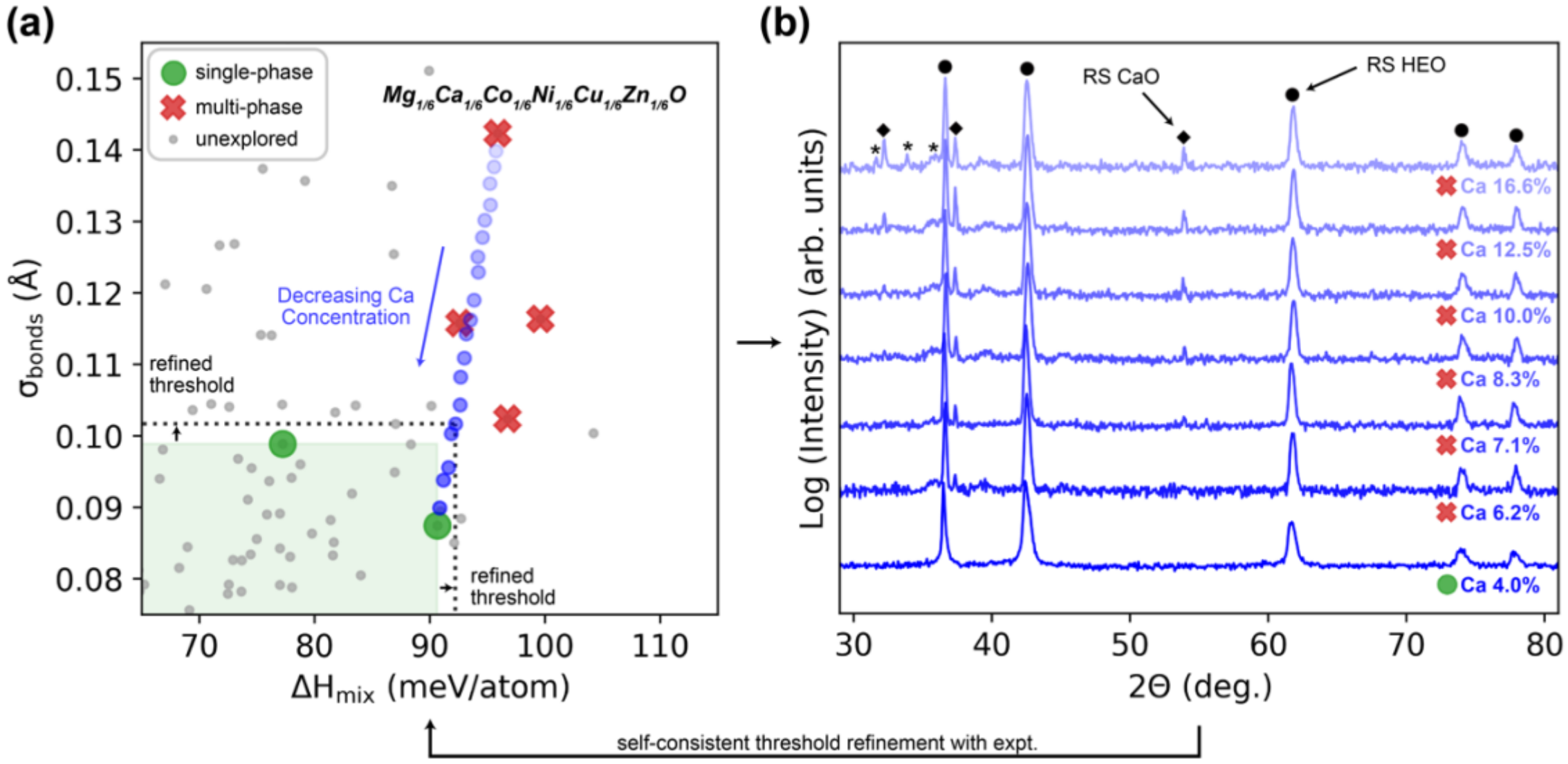

Figure 4. Exploration of non-equimolar composition space between single-phase $Mg_{1/5}Co_{1/5}Ni_{1/5}Cu_{1/5}Zn_{1/5}O$ and multi-phase $Mg_{1/6}Ca_{1/6}Co_{1/6}Ni_{1/6}Cu_{1/6}Zn_{1/6}O$ using (a) computational descriptors and (b) experimental characterization with XRD. A refined single-phase stability threshold is shown in (a) as a black dashed line.

We describe a high-throughput framework for discovering and understanding the single-phase formation of HEOs by integrating computation and experiment in a self-consistent feedback loop. To enable a more rapid exploration of rock salt HEO composition space, we use the CHGNet MLIP which exhibits impressive predictive accuracy for local structure and energies even for highly disordered systems as seen from our benchmarking. Additional fine-tuning for specific materials of interest could likely enable even greater accuracy, however we utilize the universal potential from Deng et al. to emphasize the broad applicability of pretrained MLIPs for materials spanning across the periodic table [20]. Our two computational descriptors, $\Delta H_{mix}$ and $\sigma_{bonds}$, are capable of resolving single-phase stabilities for all eight equimolar rock salt HEO compositions explored to date as well as a novel non-equimolar rock salt HEO containing Ca. Dozens of unexplored compositions are also predicted to form single-phase rock salt HEOs and await experimental discovery (Table S3, Supplemental Material [26]). Our approach currently relies on these fairly simple but highly interpretable descriptors; however, more elaborate descriptors that explicitly include the effects of configurational entropy, chemical potential, and finite temperature may be more representative of HEO stability. As more complex crystal structures such as spinel and perovskites are considered for emerging property engineering opportunities these quantities may be increasingly important. Combining such descriptors with MLIPs presents an exciting outlook to computationally drive the exploration, discovery, and implementation of HEOs and other disordered materials.

**Acknowledgements**
The authors acknowledge the use of facilities and instrumentation supported by NSF through the Pennsylvania State University Materials Research Science and Engineering Center [DMR-2011839]. Calculations utilized resources from the Roar Collab cluster of the Penn State Institute for Computational and Data Sciences.

## Supplementary Material

Table S1. Relaxed lattice parameter comparison between CHGNet and DFT for $A^{2+}O^{2-}$ ground states.

| Composition | Phase | CHGNet (Å) | DFT (Å) | Difference (Å) [%] |
|---|---|---|---|---|
| MgO | Rock salt | 4.249 | 4.258 | -0.009 [0.21%] |
| CaO | Rock salt | 4.845 | 4.840 | 0.005 [0.10%] |
| MnO | Rock salt | 4.490 | 4.517 | -0.027 [0.60%] |
| FeO | Rock salt | 4.391 | 4.321 | 0.070 [1.61%] |
| CoO | Rock salt | 4.296 | 4.253 | 0.043 [1.00%] |
| NiO | Rock salt | 4.220 | 4.229 | -0.009 [0.21%] |
| CuO | Tenorite | a = 4.249<br>b = 4.061<br>c = 5.160 | a = 4.243<br>b = 4.061<br>c = 5.151 | 0.006 [0.14%]<br>0.000 [0.00%]<br>0.009 [0.18%] |
| ZnO | Wurtzite | a = b = 3.287<br>c = 5.295 | a = b = 3.289<br>c = 5.306 | -0.002 [0.06%]<br>-0.011 [0.21%] |

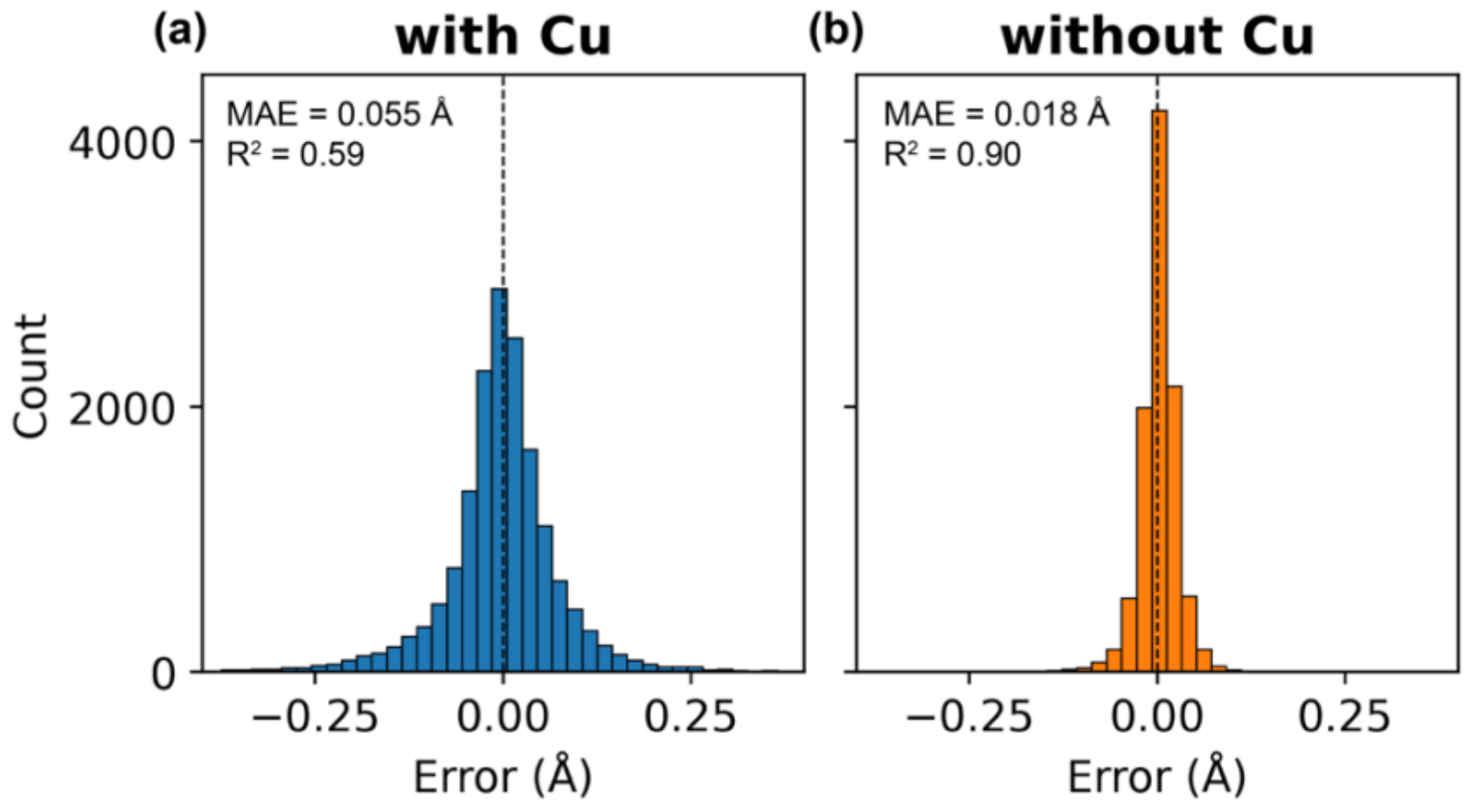

Figure S1. First near-neighbor bond length benchmarking data for CHGNet compared to DFT for compositions (a) containing Cu and (b) do not contain Cu. The x-axis is the difference between DFT and CHGNet bond lengths.

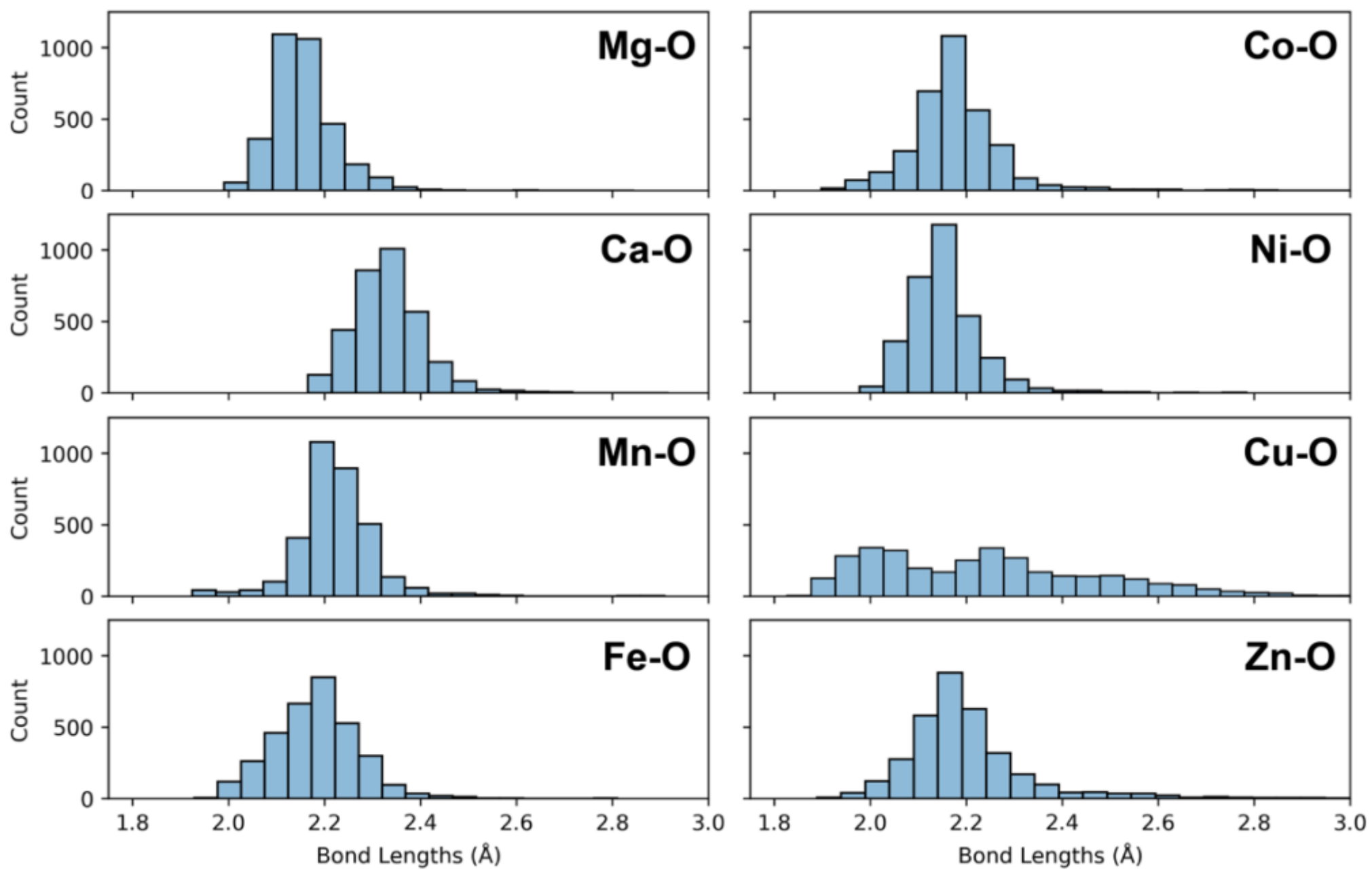

Figure S2. Distributions of CHGNet relaxed first near-neighbor bond lengths from 5-cation HEO benchmarking data for each metal-oxygen ion pair. Note the two primary peaks for Cu-O around 2.0 and 2.25 Å indicative of Jahn-Teller octahedral distortion.

Table S2. Average and standard deviation for mixing enthalpy and relaxed bond lengths for $A_{1/2}A'_{1/2}O$ rock salt oxides using CHGNet. 10 structures of 1200-atoms are used for each composition. Note that our descriptors in Figures 2a and 2b correspond to the average mixing enthalpy ($\Delta H_{mix}$) and standard deviation of bond lengths ($\sigma_{bonds}$), respectively, and are highlighted in bold for clarity.

| **Composition** | **Mixing Enthalpy (meV/atom)** | **Relaxed Bond Lengths (Å)** |
|---|---|---|
| $Mg_{1/2}Ca_{1/2}O$ | **86.06** ± 0.63 | 2.290 ± **0.093** |
| $Mg_{1/2}Mn_{1/2}O$ | **39.57** ± 0.59 | 2.192 ± **0.045** |
| $Mg_{1/2}Fe_{1/2}O$ | **42.56** ± 0.42 | 2.161 ± **0.039** |
| $Mg_{1/2}Co_{1/2}O$ | **61.39** ± 0.90 | 2.133 ± **0.018** |
| $Mg_{1/2}Ni_{1/2}O$ | **34.39** ± 0.71 | 2.113 ± **0.010** |
| $Mg_{1/2}Cu_{1/2}O$ | **71.12** ± 0.86 | 2.154 ± **0.166** |
| $Mg_{1/2}Zn_{1/2}O$ | **52.99** ± 0.14 | 2.144 ± **0.022** |
| $Ca_{1/2}Mn_{1/2}O$ | **36.64** ± 0.17 | 2.337 ± **0.053** |
| $Ca_{1/2}Fe_{1/2}O$ | **66.14** ± 0.22 | 2.317 ± **0.108** |
| $Ca_{1/2}Co_{1/2}O$ | **90.40** ± 0.43 | 2.296 ± **0.141** |
| $Ca_{1/2}Ni_{1/2}O$ | **108.48** ± 0.55 | 2.277 ± **0.107** |
| $Ca_{1/2}Cu_{1/2}O$ | **96.79** ± 0.98 | 2.350 ± **0.310** |
| $Ca_{1/2}Zn_{1/2}O$ | **96.35** ± 0.78 | 2.329 ± **0.203** |
| $Mn_{1/2}Fe_{1/2}O$ | **39.86** ± 0.03 | 2.230 ± **0.036** |
| $Mn_{1/2}Co_{1/2}O$ | **47.94** ± 0.12 | 2.189 ± **0.031** |
| $Mn_{1/2}Ni_{1/2}O$ | **37.89** ± 0.09 | 2.179 ± **0.044** |
| $Mn_{1/2}Cu_{1/2}O$ | **35.08** ± 0.73 | 2.230 ± **0.233** |
| $Mn_{1/2}Zn_{1/2}O$ | **97.98** ± 0.09 | 2.212 ± **0.036** |
| $Fe_{1/2}Co_{1/2}O$ | **73.53** ± 0.05 | 2.158 ± **0.016** |
| $Fe_{1/2}Ni_{1/2}O$ | **69.04** ± 0.12 | 2.144 ± **0.033** |
| $Fe_{1/2}Cu_{1/2}O$ | **56.25** ± 0.84 | 2.200 ± **0.212** |
| $Fe_{1/2}Zn_{1/2}O$ | **106.89** ± 0.05 | 2.178 ± **0.017** |
| $Co_{1/2}Ni_{1/2}O$ | **69.73** ± 0.04 | 2.122 ± **0.011** |
| $Co_{1/2}Cu_{1/2}O$ | **69.56** ± 0.46 | 2.169 ± **0.193** |
| $Co_{1/2}Zn_{1/2}O$ | **113.33** ± 0.09 | 2.154 ± **0.012** |
| $Ni_{1/2}Cu_{1/2}O$ | **84.08** ± 0.57 | 2.143 ± **0.165** |
| $Ni_{1/2}Zn_{1/2}O$ | **110.36** ± 0.08 | 2.136 ± **0.020** |
| $Cu_{1/2}Zn_{1/2}O$ | **149.00** ± 1.22 | 2.218 ± **0.281** |

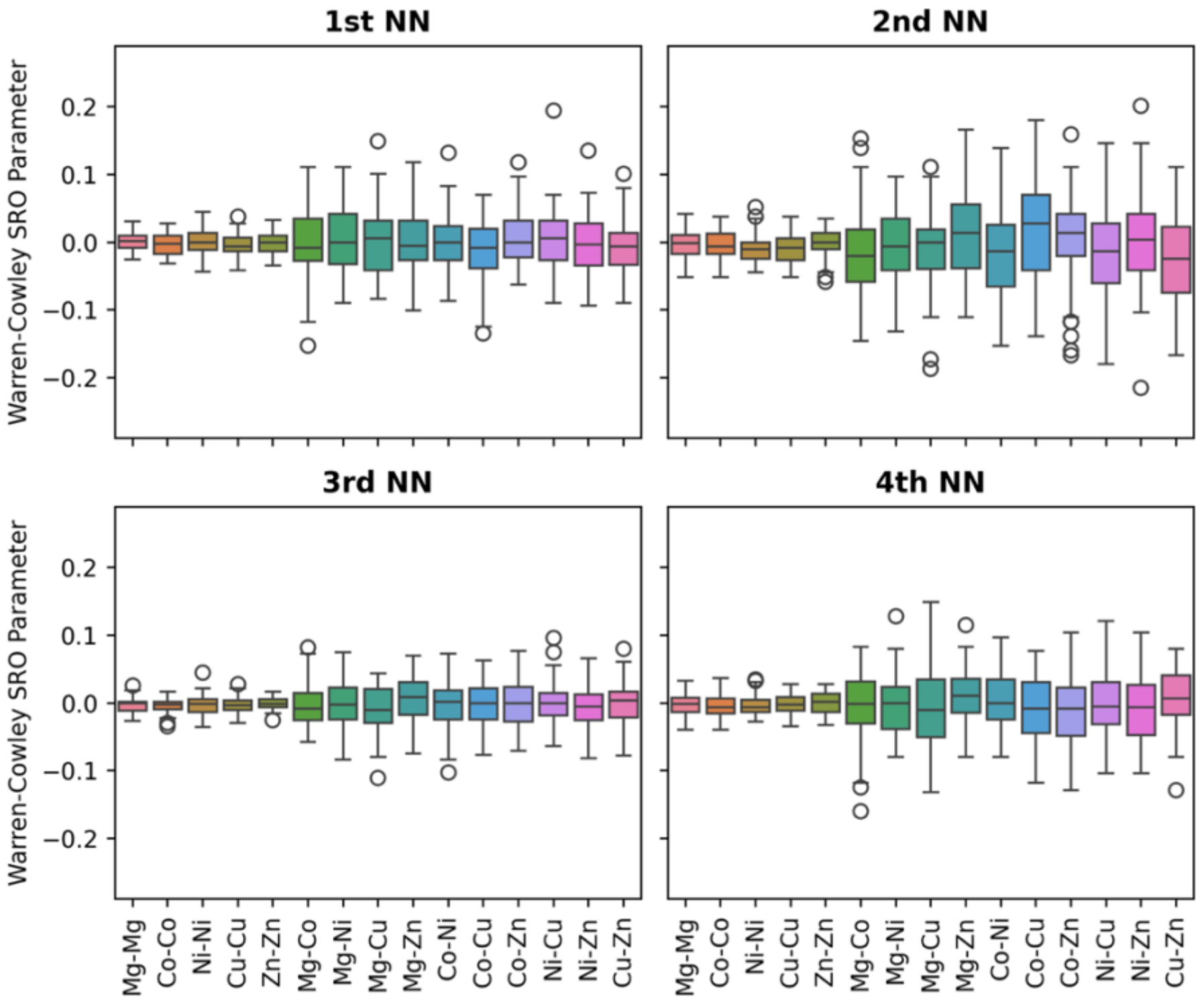

Figure S3. Warren-Cowley short-range order (SRO) parameters for 50 sampled $Mg_{1/5}Co_{1/5}Ni_{1/5}Cu_{1/5}Zn_{1/5}O$ randomly decorated structures for 1$^{st}$, 2$^{nd}$, 3$^{rd}$, and 4$^{th}$ near-neighbor (NN) cation pairs. For clarity, we only show SRO pair parameters for a single HEO composition, the prototypical HEO $Mg_{1/5}Co_{1/5}Ni_{1/5}Cu_{1/5}Zn_{1/5}O$.

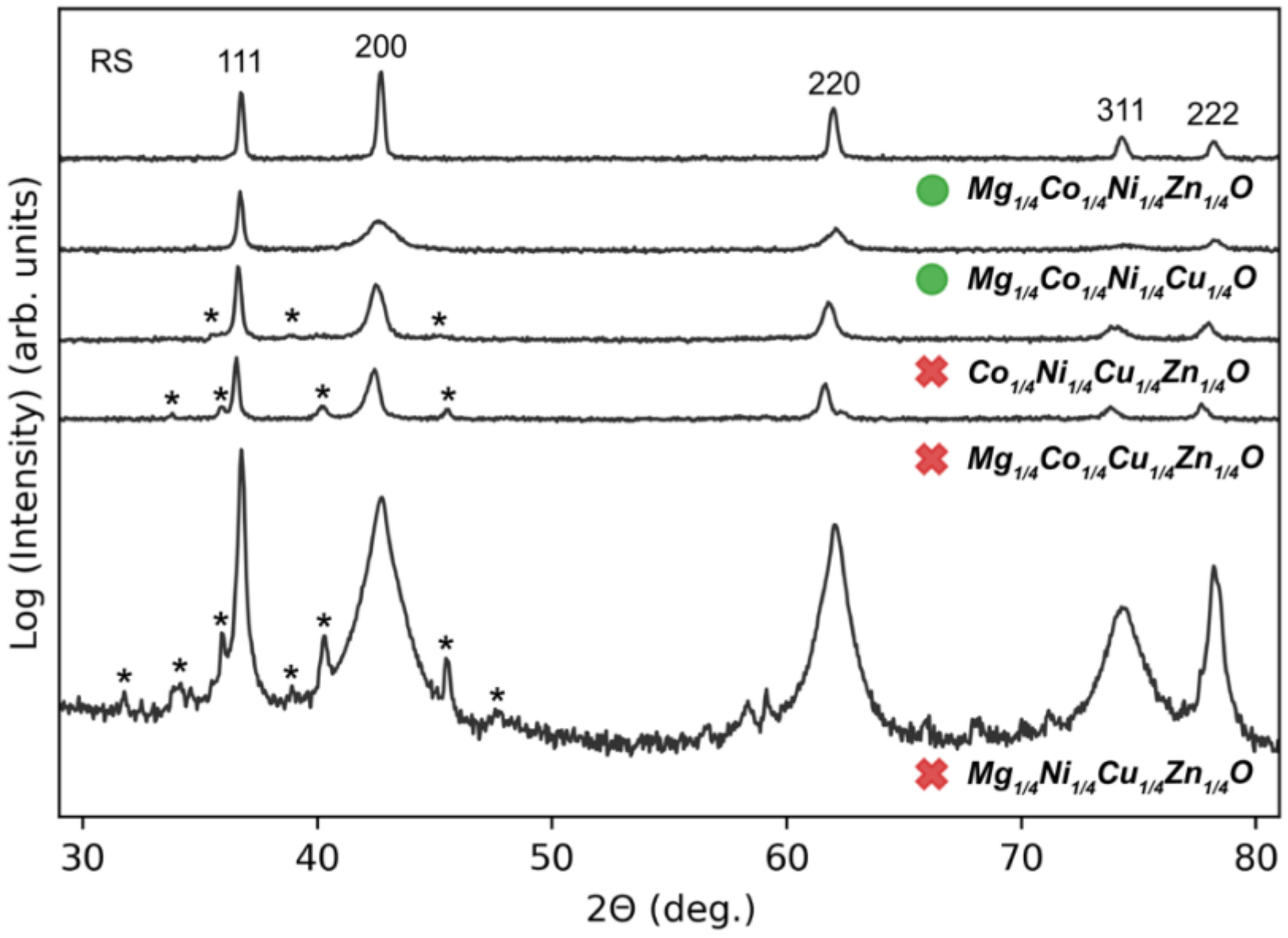

Figure S4. X-ray diffraction patterns for five four-cation derivatives of $Mg_{1/5}Co_{1/5}Ni_{1/5}Cu_{1/5}Zn_{1/5}O$. Experimental synthesis results for single- and multi-phase are indicated with green circle and red crosses, respectively.

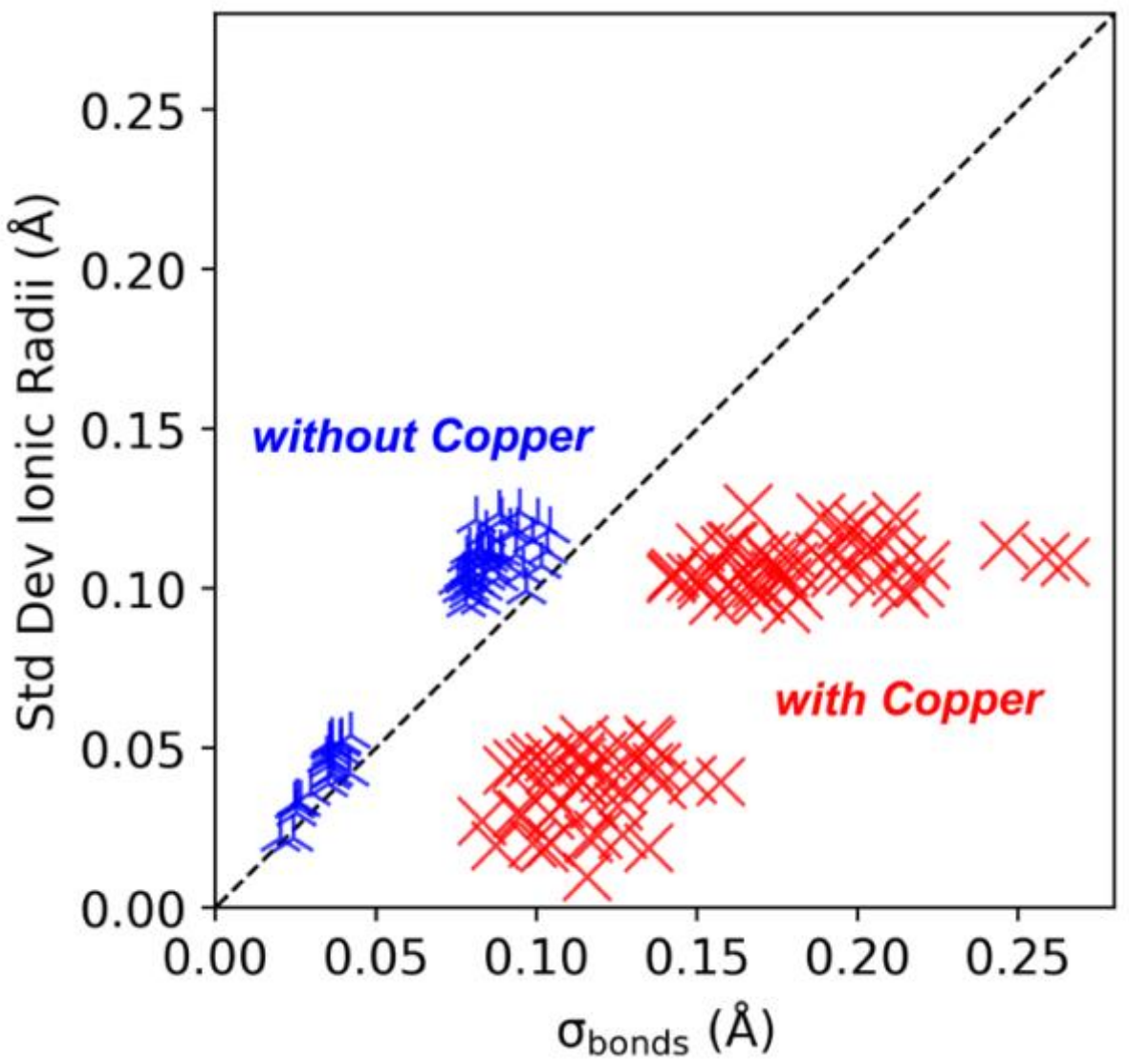

Figure S5. Comparison between standard deviation of CHGNet calculated first near-neighbor bond lengths ($\sigma_{bonds}$) and standard deviation of individual Shannon ionic radii. Data is separated between compositions that contain Cu due to the known Jahn-Teller distortion of local octahedra.

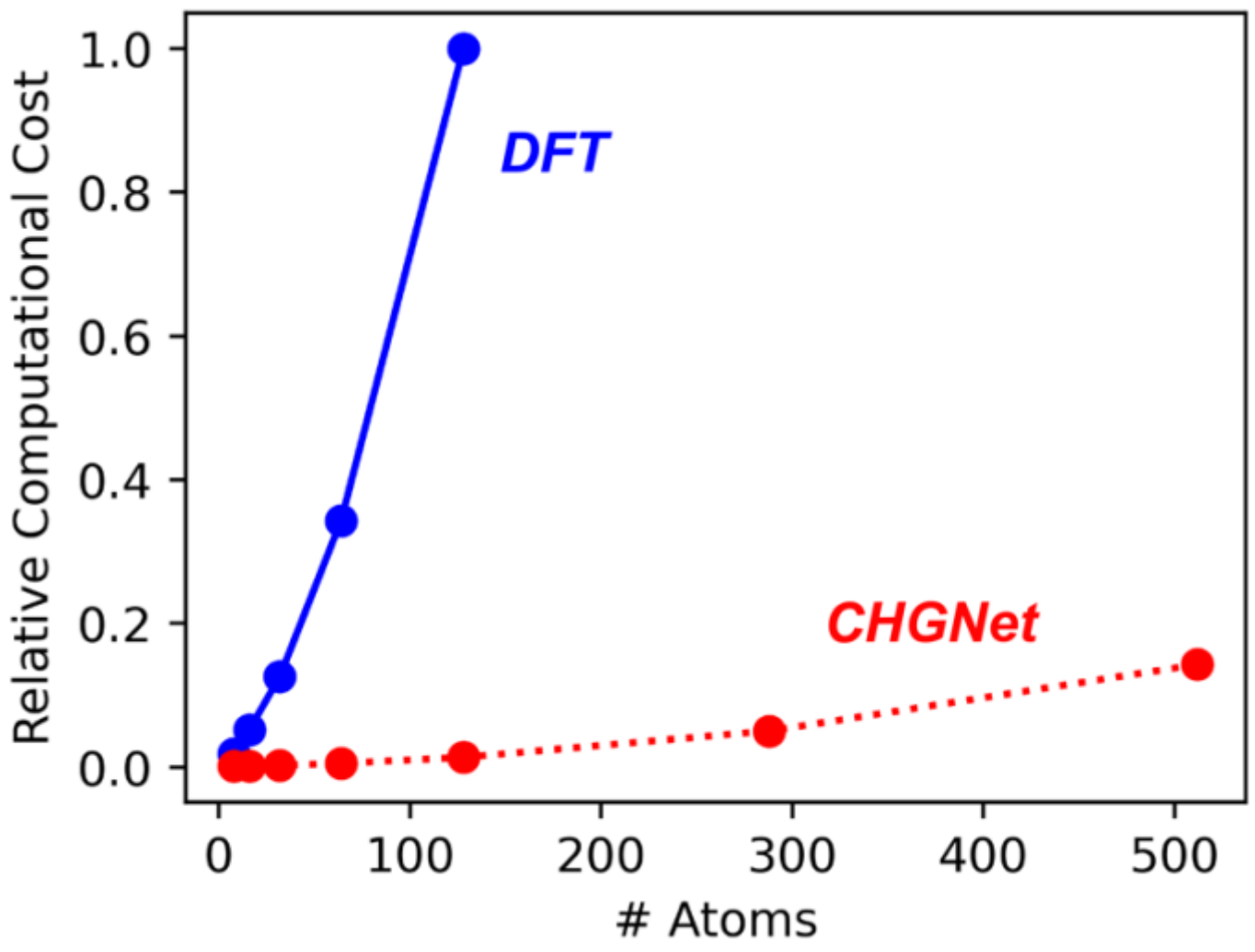

Figure S6. Comparison of relative computation cost between VASP DFT calculations and CHGNet predictions to qualitatively demonstrate the large reduction of computational costs afforded by MLIPs. Various supercells sizes of MgO were utilized for consistency between runs.

Table S3. Average and standard deviation for mixing enthalpy and relaxed bond lengths for equimolar HEO compositions from cation cohort: Mg, Ca, Mn, Fe, Co, Ni, Cu, and Zn. Note that our descriptors plotted in Figure 3 correspond to the average mixing enthalpy ($\Delta H_{mix}$) and standard deviation of bond lengths ($\sigma_{bonds}$) and are highlighted in bold for clarity. Compositions are sorted by $\sigma_{bonds}$. The refined single-phase stability thresholds from Figure 4 are used to determine predicted single-phase stability: $\Delta H_{mix}$ = 92.2 meV/atom, $\sigma_{bonds}$ = 0.102 Å. Predicted and experimental stability as single- and multi-phase are indicated with green circle and red crosses, respectively. Compositions not yet explored experimentally are indicated by two dashes.

| Composition | Relaxed Bond Lengths (Å) | Mixing Enthalpy (meV/atom) | Pred. Stability | Expt. Stability |
|---|---|---|---|---|
| $Mg_{1/4}Co_{1/4}Ni_{1/4}Zn_{1/4}O$ | 2.133 ± **0.020** | **78.68** ± 0.50 | ● | ● |
| $Mg_{1/4}Fe_{1/4}Co_{1/4}Zn_{1/4}O$ | 2.156 ± **0.024** | **75.27** ± 0.53 | ● | -- |
| $Fe_{1/4}Co_{1/4}Ni_{1/4}Zn_{1/4}O$ | 2.149 ± **0.025** | **89.64** ± 0.13 | ● | -- |
| $Mg_{1/5}Fe_{1/5}Co_{1/5}Ni_{1/5}Zn_{1/5}O$ | 2.146 ± **0.025** | **77.21** ± 0.48 | ● | -- |
| $Mg_{1/4}Fe_{1/4}Co_{1/4}Ni_{1/4}O$ | 2.140 ± **0.026** | **70.65** ± 0.52 | ● | -- |
| $Mg_{1/4}Fe_{1/4}Ni_{1/4}Zn_{1/4}O$ | 2.147 ± **0.027** | **67.57** ± 0.35 | ● | -- |
| $Mn_{1/4}Fe_{1/4}Co_{1/4}Zn_{1/4}O$ | 2.186 ± **0.030** | **77.51** ± 0.11 | ● | -- |
| $Mn_{1/5}Fe_{1/5}Co_{1/5}Ni_{1/5}Zn_{1/5}O$ | 2.171 ± **0.035** | **76.31** ± 0.15 | ● | -- |
| $Mg_{1/5}Mn_{1/5}Fe_{1/5}Co_{1/5}Zn_{1/5}O$ | 2.176 ± **0.035** | **70.65** ± 0.36 | ● | -- |
| $Mg_{1/6}Mn_{1/6}Fe_{1/6}Co_{1/6}Ni_{1/6}Zn_{1/6}O$ | 2.165 ± **0.036** | **71.21** ± 0.38 | ● | -- |
| $Mn_{1/4}Fe_{1/4}Co_{1/4}Ni_{1/4}O$ | 2.169 ± **0.036** | **58.84** ± 0.11 | ● | -- |
| $Mg_{1/5}Mn_{1/5}Co_{1/5}Ni_{1/5}Zn_{1/5}O$ | 2.158 ± **0.036** | **71.04** ± 0.37 | ● | -- |
| $Mg_{1/4}Mn_{1/4}Fe_{1/4}Co_{1/4}O$ | 2.178 ± **0.036** | **57.34** ± 0.46 | ● | -- |
| $Mn_{1/4}Co_{1/4}Ni_{1/4}Zn_{1/4}O$ | 2.166 ± **0.036** | **79.18** ± 0.12 | ● | -- |
| $Mg_{1/5}Mn_{1/5}Fe_{1/5}Co_{1/5}Ni_{1/5}O$ | 2.164 ± **0.036** | **61.63** ± 0.42 | ● | ● [13] |
| $Mg_{1/4}Mn_{1/4}Co_{1/4}Ni_{1/4}O$ | 2.156 ± **0.036** | **59.86** ± 0.41 | ● | -- |
| $Mg_{1/4}Mn_{1/4}Co_{1/4}Zn_{1/4}O$ | 2.171 ± **0.037** | **71.57** ± 0.47 | ● | -- |
| $Mn_{1/4}Fe_{1/4}Ni_{1/4}Zn_{1/4}O$ | 2.182 ± **0.039** | **75.68** ± 0.18 | ● | -- |
| $Mg_{1/5}Mn_{1/5}Fe_{1/5}Ni_{1/5}Zn_{1/5}O$ | 2.171 ± **0.039** | **66.34** ± 0.38 | ● | -- |
| $Mg_{1/4}Mn_{1/4}Ni_{1/4}Zn_{1/4}O$ | 2.163 ± **0.039** | **61.32** ± 0.38 | ● | -- |
| $Mg_{1/4}Mn_{1/4}Fe_{1/4}Zn_{1/4}O$ | 2.187 ± **0.042** | **70.35** ± 0.40 | ● | -- |
| $Mg_{1/4}Mn_{1/4}Fe_{1/4}Ni_{1/4}O$ | 2.173 ± **0.042** | **52.87** ± 0.53 | ● | -- |
| $Mg_{1/6}Ca_{1/6}Mn_{1/6}Fe_{1/6}Co_{1/6}Ni_{1/6}O$ | 2.217 ± **0.076** | **69.10** ± 0.46 | ● | -- |
| $Mg_{1/5}Ca_{1/5}Mn_{1/5}Fe_{1/5}Co_{1/5}O$ | 2.236 ± **0.077** | **64.52** ± 0.47 | ● | -- |
| $Mg_{1/4}Ca_{1/4}Mn_{1/4}Fe_{1/4}O$ | 2.259 ± **0.077** | **63.51** ± 0.53 | ● | -- |
| $Mg_{1/6}Ca_{1/6}Mn_{1/6}Fe_{1/6}Ni_{1/6}Zn_{1/6}O$ | 2.219 ± **0.078** | **72.47** ± 0.46 | ● | -- |
| $Mg_{1/6}Ca_{1/6}Mn_{1/6}Fe_{1/6}Co_{1/6}Zn_{1/6}O$ | 2.226 ± **0.078** | **73.62** ± 0.46 | ● | -- |
| $Mg_{1/5}Ca_{1/5}Mn_{1/5}Fe_{1/5}Ni_{1/5}O$ | 2.229 ± **0.078** | **64.79** ± 0.51 | ● | -- |
| $Mg_{1/6}Ca_{1/6}Mn_{1/6}Co_{1/6}Ni_{1/6}Zn_{1/6}O$ | 2.212 ± **0.079** | **78.03** ± 0.63 | ● | -- |
| $Ca_{1/4}Mn_{1/4}Fe_{1/4}Co_{1/4}O$ | 2.257 ± **0.079** | **55.93** ± 0.36 | ● | -- |
| $Ca_{1/6}Mn_{1/6}Fe_{1/6}Co_{1/6}Ni_{1/6}Zn_{1/6}O$ | 2.223 ± **0.079** | **76.97** ± 0.47 | ● | -- |
| $Mg_{1/5}Ca_{1/5}Mn_{1/5}Co_{1/5}Ni_{1/5}O$ | 2.220 ± **0.079** | **72.52** ± 0.52 | ● | -- |
| $Ca_{1/5}Mn_{1/5}Fe_{1/5}Co_{1/5}Ni_{1/5}O$ | 2.230 ± **0.079** | **65.20** ± 0.44 | ● | -- |
| $Mg_{1/6}Ca_{1/6}Fe_{1/6}Co_{1/6}Ni_{1/6}Zn_{1/6}O$ | 2.202 ± **0.080** | **84.00** ± 0.50 | ● | -- |
| $Mg_{1/4}Ca_{1/4}Mn_{1/4}Ni_{1/4}O$ | 2.237 ± **0.081** | **64.69** ± 0.64 | ● | -- |
| $Mg_{1/4}Ca_{1/4}Mn_{1/4}Co_{1/4}O$ | 2.245 ± **0.082** | **68.21** ± 0.59 | ● | -- |
| $Ca_{1/4}Mn_{1/4}Fe_{1/4}Ni_{1/4}O$ | 2.254 ± **0.082** | **60.25** ± 0.47 | ● | -- |
| $Ca_{1/5}Mn_{1/5}Fe_{1/5}Co_{1/5}Zn_{1/5}O$ | 2.244 ± **0.083** | **73.67** ± 0.47 | ● | -- |
| $Mg_{1/5}Ca_{1/5}Mn_{1/5}Ni_{1/5}Zn_{1/5}O$ | 2.223 ± **0.083** | **72.91** ± 0.68 | ● | -- |
| $Mg_{1/6}Fe_{1/6}Co_{1/6}Ni_{1/6}Cu_{1/6}Zn_{1/6}O$ | 2.152 ± **0.083** | **77.88** ± 0.51 | ● | -- |
| $Mg_{1/5}Ca_{1/5}Fe_{1/5}Co_{1/5}Ni_{1/5}O$ | 2.208 ± **0.083** | **81.57** ± 0.58 | ● | -- |
| $Mg_{1/5}Ca_{1/5}Mn_{1/5}Fe_{1/5}Zn_{1/5}O$ | 2.243 ± **0.083** | **74.44** ± 0.56 | ● | -- |
| $Mg_{1/5}Ca_{1/5}Mn_{1/5}Co_{1/5}Zn_{1/5}O$ | 2.231 ± **0.084** | **76.96** ± 0.47 | ● | -- |
| $Ca_{1/4}Mn_{1/4}Co_{1/4}Ni_{1/4}O$ | 2.237 ± **0.084** | **68.95** ± 0.51 | ● | -- |
| $Mg_{1/5}Ca_{1/5}Co_{1/5}Ni_{1/5}Zn_{1/5}O$ | 2.201 ± **0.085** | **92.11** ± 0.75 | ● | -- |
| $Ca_{1/5}Mn_{1/5}Co_{1/5}Ni_{1/5}Zn_{1/5}O$ | 2.228 ± **0.085** | **81.66** ± 0.56 | ● | -- |
| $Ca_{1/5}Mn_{1/5}Fe_{1/5}Ni_{1/5}Zn_{1/5}O$ | 2.240 ± **0.086** | **74.98** ± 0.48 | ● | -- |
| $Mg_{1/5}Ca_{1/5}Fe_{1/5}Ni_{1/5}Zn_{1/5}O$ | 2.212 ± **0.086** | **79.76** ± 0.66 | ● | -- |
| $Mg_{1/5}Co_{1/5}Ni_{1/5}Cu_{1/5}Zn_{1/5}O$ | 2.143 ± **0.087** | **90.63** ± 0.53 | ● | ● [3] |
| $Mg_{1/5}Ca_{1/5}Fe_{1/5}Co_{1/5}Zn_{1/5}O$ | 2.220 ± **0.088** | **81.38** ± 0.62 | ● | -- |
| $Mg_{1/4}Ca_{1/4}Co_{1/4}Ni_{1/4}O$ | 2.210 ± **0.088** | **92.71** ± 0.70 | ✖ | -- |
| $Mg_{1/4}Ca_{1/4}Fe_{1/4}Co_{1/4}O$ | 2.231 ± **0.089** | **75.83** ± 0.60 | ● | -- |
| $Mg_{1/4}Ca_{1/4}Fe_{1/4}Ni_{1/4}O$ | 2.222 ± **0.089** | **76.99** ± 0.75 | ● | -- |
| $Ca_{1/5}Fe_{1/5}Co_{1/5}Ni_{1/5}Zn_{1/5}O$ | 2.216 ± **0.089** | **90.91** ± 0.53 | ● | -- |
| $Mg_{1/6}Mn_{1/6}Co_{1/6}Ni_{1/6}Cu_{1/6}Zn_{1/6}O$ | 2.163 ± **0.091** | **74.18** ± 0.54 | ● | -- |
| $Ca_{1/4}Fe_{1/4}Co_{1/4}Ni_{1/4}O$ | 2.224 ± **0.092** | **83.24** ± 0.58 | ● | -- |
| $Mg_{1/4}Ca_{1/4}Mn_{1/4}Zn_{1/4}O$ | 2.254 ± **0.094** | **76.05** ± 0.62 | ● | -- |
| $Mg_{1/5}Fe_{1/5}Co_{1/5}Ni_{1/5}Cu_{1/5}O$ | 2.150 ± **0.094** | **66.52** ± 0.45 | ● | -- |
| $Mg_{1/6}Mn_{1/6}Fe_{1/6}Co_{1/6}Ni_{1/6}Cu_{1/6}O$ | 2.168 ± **0.094** | **57.74** ± 0.34 | ● | -- |
| $Ca_{1/4}Mn_{1/4}Ni_{1/4}Zn_{1/4}O$ | 2.250 ± **0.094** | **77.98** ± 0.62 | ● | -- |
| $Mg_{1/4}Ca_{1/4}Ni_{1/4}Zn_{1/4}O$ | 2.216 ± **0.095** | **86.94** ± 0.52 | ● | -- |
| $Mg_{1/5}Fe_{1/5}Ni_{1/5}Cu_{1/5}Zn_{1/5}O$ | 2.155 ± **0.096** | **74.50** ± 0.59 | ● | -- |
| $Ca_{1/4}Mn_{1/4}Co_{1/4}Zn_{1/4}O$ | 2.255 ± **0.096** | **78.74** ± 0.45 | ● | -- |
| $Ca_{1/4}Mn_{1/4}Fe_{1/4}Zn_{1/4}O$ | 2.270 ± **0.097** | **73.33** ± 0.45 | ● | -- |
| $Mg_{1/6}Mn_{1/6}Fe_{1/6}Ni_{1/6}Cu_{1/6}Zn_{1/6}O$ | 2.173 ± **0.098** | **66.80** ± 0.54 | ● | -- |
| $Mg_{1/4}Ca_{1/4}Co_{1/4}Zn_{1/4}O$ | 2.226 ± **0.099** | **88.33** ± 0.71 | ● | -- |
| $Mg_{1/4}Co_{1/4}Ni_{1/4}Cu_{1/4}O$ | 2.136 ± **0.099** | **77.23** ± 0.56 | ● | ● |
| $Mg_{1/5}Mn_{1/5}Co_{1/5}Ni_{1/5}Cu_{1/5}O$ | 2.161 ± **0.100** | **59.56** ± 0.52 | ● | -- |
| $Ca_{1/4}Co_{1/4}Ni_{1/4}Zn_{1/4}O$ | 2.221 ± **0.100** | **104.19** ± 0.73 | ✖ | -- |
| $Ca_{1/4}Fe_{1/4}Co_{1/4}Zn_{1/4}O$ | 2.242 ± **0.102** | **87.00** ± 0.55 | ● | -- |
| $Mg_{1/4}Ni_{1/4}Cu_{1/4}Zn_{1/4}O$ | 2.144 ± **0.102** | **96.70** ± 0.53 | ✖ | ✖ |
| $Mg_{1/4}Ca_{1/4}Fe_{1/4}Zn_{1/4}O$ | 2.240 ± **0.103** | **81.79** ± 0.61 | ✖ | -- |
| $Mg_{1/6}Mn_{1/6}Fe_{1/6}Co_{1/6}Cu_{1/6}Zn_{1/6}O$ | 2.181 ± **0.104** | **69.38** ± 0.48 | ✖ | -- |
| $Mg_{1/5}Mn_{1/5}Ni_{1/5}Cu_{1/5}Zn_{1/5}O$ | 2.166 ± **0.104** | **72.57** ± 0.54 | ✖ | -- |
| $Ca_{1/4}Fe_{1/4}Ni_{1/4}Zn_{1/4}O$ | 2.236 ± **0.104** | **90.10** ± 0.78 | ✖ | -- |
| $Fe_{1/5}Co_{1/5}Ni_{1/5}Cu_{1/5}Zn_{1/5}O$ | 2.161 ± **0.104** | **83.53** ± 0.35 | ✖ | -- |
| $Mg_{1/5}Fe_{1/5}Co_{1/5}Cu_{1/5}Zn_{1/5}O$ | 2.165 ± **0.104** | **77.17** ± 0.55 | ✖ | -- |
| $Mn_{1/6}Fe_{1/6}Co_{1/6}Ni_{1/6}Cu_{1/6}Zn_{1/6}O$ | 2.177 ± **0.104** | **70.99** ± 0.41 | ✖ | -- |
| $Mg_{1/4}Fe_{1/4}Ni_{1/4}Cu_{1/4}O$ | 2.152 ± **0.107** | **59.79** ± 0.60 | ✖ | -- |
| $Mg_{1/5}Mn_{1/5}Fe_{1/5}Ni_{1/5}Cu_{1/5}O$ | 2.174 ± **0.110** | **51.57** ± 0.55 | ✖ | -- |
| $Mg_{1/5}Mn_{1/5}Fe_{1/5}Co_{1/5}Cu_{1/5}O$ | 2.181 ± **0.114** | **54.61** ± 0.62 | ✖ | -- |
| $Mn_{1/5}Fe_{1/5}Co_{1/5}Ni_{1/5}Cu_{1/5}O$ | 2.177 ± **0.114** | **51.49** ± 0.33 | ✖ | -- |
| $Mn_{1/5}Co_{1/5}Ni_{1/5}Cu_{1/5}Zn_{1/5}O$ | 2.177 ± **0.114** | **76.24** ± 0.39 | ✖ | -- |
| $Mg_{1/5}Mn_{1/5}Co_{1/5}Cu_{1/5}Zn_{1/5}O$ | 2.174 ± **0.114** | **75.32** ± 0.65 | ✖ | -- |
| $Mg_{1/4}Mn_{1/4}Ni_{1/4}Cu_{1/4}O$ | 2.164 ± **0.115** | **53.06** ± 0.67 | ✖ | -- |
| $Mg_{1/4}Co_{1/4}Cu_{1/4}Zn_{1/4}O$ | 2.157 ± **0.116** | **92.59** ± 0.60 | ✖ | ✖ |
| $Co_{1/4}Ni_{1/4}Cu_{1/4}Zn_{1/4}O$ | 2.153 ± **0.116** | **99.58** ± 0.42 | ✖ | ✖ |
| $Fe_{1/4}Co_{1/4}Ni_{1/4}Cu_{1/4}O$ | 2.157 ± **0.118** | **62.26** ± 0.32 | ✖ | -- |
| $Mg_{1/4}Fe_{1/4}Co_{1/4}Cu_{1/4}O$ | 2.164 ± **0.119** | **62.64** ± 0.75 | ✖ | -- |
| $Mn_{1/5}Fe_{1/5}Ni_{1/5}Cu_{1/5}Zn_{1/5}O$ | 2.185 ± **0.121** | **70.60** ± 0.44 | ✖ | -- |
| $Mg_{1/5}Mn_{1/5}Fe_{1/5}Cu_{1/5}Zn_{1/5}O$ | 2.188 ± **0.121** | **67.02** ± 0.67 | ✖ | -- |
| $Fe_{1/4}Ni_{1/4}Cu_{1/4}Zn_{1/4}O$ | 2.167 ± **0.125** | **86.84** ± 0.56 | ✖ | -- |
| $Mn_{1/4}Co_{1/4}Ni_{1/4}Cu_{1/4}O$ | 2.172 ± **0.126** | **49.96** ± 0.42 | ✖ | -- |
| $Mn_{1/5}Fe_{1/5}Co_{1/5}Cu_{1/5}Zn_{1/5}O$ | 2.193 ± **0.127** | **71.72** ± 0.40 | ✖ | -- |
| $Mg_{1/4}Fe_{1/4}Cu_{1/4}Zn_{1/4}O$ | 2.171 ± **0.127** | **73.01** ± 0.72 | ✖ | -- |
| $Mg_{1/4}Mn_{1/4}Co_{1/4}Cu_{1/4}O$ | 2.178 ± **0.128** | **56.89** ± 0.60 | ✖ | -- |
| $Mn_{1/4}Fe_{1/4}Ni_{1/4}Cu_{1/4}O$ | 2.186 ± **0.135** | **45.49** ± 0.49 | ✖ | -- |
| $Fe_{1/4}Co_{1/4}Cu_{1/4}Zn_{1/4}O$ | 2.178 ± **0.135** | **86.70** ± 0.45 | ✖ | -- |
| $Mg_{1/4}Mn_{1/4}Fe_{1/4}Cu_{1/4}O$ | 2.191 ± **0.135** | **47.61** ± 0.64 | ✖ | -- |
| $Mn_{1/4}Ni_{1/4}Cu_{1/4}Zn_{1/4}O$ | 2.181 ± **0.136** | **79.14** ± 0.51 | ✖ | -- |
| $Mg_{1/4}Mn_{1/4}Cu_{1/4}Zn_{1/4}O$ | 2.185 ± **0.137** | **75.47** ± 0.63 | ✖ | -- |
| $Mn_{1/4}Fe_{1/4}Co_{1/4}Cu_{1/4}O$ | 2.196 ± **0.139** | **47.34** ± 0.51 | ✖ | -- |
| $Mg_{1/6}Ca_{1/6}Fe_{1/6}Co_{1/6}Ni_{1/6}Cu_{1/6}O$ | 2.208 ± **0.142** | **73.76** ± 0.69 | ✖ | -- |
| $Mg_{1/6}Ca_{1/6}Co_{1/6}Ni_{1/6}Cu_{1/6}Zn_{1/6}O$ | 2.204 ± **0.142** | **95.89** ± 0.73 | ✖ | ✖ |
| $Mg_{1/6}Ca_{1/6}Mn_{1/6}Co_{1/6}Ni_{1/6}Cu_{1/6}O$ | 2.216 ± **0.143** | **68.86** ± 0.69 | ✖ | -- |
| $Mg_{1/6}Ca_{1/6}Mn_{1/6}Fe_{1/6}Ni_{1/6}Cu_{1/6}O$ | 2.226 ± **0.148** | **58.88** ± 0.85 | ✖ | -- |
| $Mn_{1/4}Co_{1/4}Cu_{1/4}Zn_{1/4}O$ | 2.194 ± **0.149** | **79.51** ± 0.49 | ✖ | -- |
| $Mg_{1/5}Ca_{1/5}Co_{1/5}Ni_{1/5}Cu_{1/5}O$ | 2.209 ± **0.151** | **89.93** ± 0.84 | ✖ | -- |
| $Mg_{1/6}Ca_{1/6}Fe_{1/6}Ni_{1/6}Cu_{1/6}Zn_{1/6}O$ | 2.215 ± **0.151** | **79.12** ± 0.81 | ✖ | -- |
| $Ca_{1/6}Mn_{1/6}Fe_{1/6}Co_{1/6}Ni_{1/6}Cu_{1/6}O$ | 2.227 ± **0.151** | **56.26** ± 0.77 | ✖ | -- |
| $Mg_{1/6}Ca_{1/6}Mn_{1/6}Fe_{1/6}Co_{1/6}Cu_{1/6}O$ | 2.233 ± **0.155** | **57.67** ± 0.86 | ✖ | -- |
| $Mg_{1/6}Ca_{1/6}Mn_{1/6}Ni_{1/6}Cu_{1/6}Zn_{1/6}O$ | 2.223 ± **0.156** | **77.66** ± 0.79 | ✖ | -- |
| $Mn_{1/4}Fe_{1/4}Cu_{1/4}Zn_{1/4}O$ | 2.209 ± **0.157** | **73.13** ± 0.69 | ✖ | -- |
| $Mg_{1/5}Ca_{1/5}Fe_{1/5}Ni_{1/5}Cu_{1/5}O$ | 2.221 ± **0.159** | **71.15** ± 1.16 | ✖ | -- |
| $Ca_{1/6}Fe_{1/6}Co_{1/6}Ni_{1/6}Cu_{1/6}Zn_{1/6}O$ | 2.219 ± **0.160** | **82.50** ± 0.92 | ✖ | -- |
| $Mg_{1/5}Ca_{1/5}Ni_{1/5}Cu_{1/5}Zn_{1/5}O$ | 2.217 ± **0.161** | **102.68** ± 1.05 | ✖ | -- |
| $Mg_{1/5}Ca_{1/5}Mn_{1/5}Ni_{1/5}Cu_{1/5}O$ | 2.230 ± **0.161** | **66.49** ± 0.96 | ✖ | -- |
| $Mg_{1/6}Ca_{1/6}Fe_{1/6}Co_{1/6}Cu_{1/6}Zn_{1/6}O$ | 2.223 ± **0.163** | **77.25** ± 0.86 | ✖ | -- |
| $Ca_{1/6}Mn_{1/6}Co_{1/6}Ni_{1/6}Cu_{1/6}Zn_{1/6}O$ | 2.228 ± **0.165** | **76.99** ± 0.76 | ✖ | -- |
| $Mg_{1/4}Ca_{1/4}Ni_{1/4}Cu_{1/4}O$ | 2.224 ± **0.166** | **98.23** ± 0.90 | ✖ | -- |
| $Ca_{1/5}Fe_{1/5}Co_{1/5}Ni_{1/5}Cu_{1/5}O$ | 2.223 ± **0.168** | **68.88** ± 0.74 | ✖ | -- |
| $Mg_{1/6}Ca_{1/6}Mn_{1/6}Co_{1/6}Cu_{1/6}Zn_{1/6}O$ | 2.232 ± **0.168** | **75.92** ± 0.93 | ✖ | -- |
| $Ca_{1/6}Mn_{1/6}Fe_{1/6}Ni_{1/6}Cu_{1/6}Zn_{1/6}O$ | 2.238 ± **0.169** | **68.42** ± 0.98 | ✖ | -- |
| $Mg_{1/6}Ca_{1/6}Mn_{1/6}Fe_{1/6}Cu_{1/6}Zn_{1/6}O$ | 2.241 ± **0.170** | **66.61** ± 0.96 | ✖ | -- |
| $Ca_{1/5}Mn_{1/5}Co_{1/5}Ni_{1/5}Cu_{1/5}O$ | 2.234 ± **0.171** | **60.34** ± 0.90 | ✖ | -- |
| $Ca_{1/5}Mn_{1/5}Fe_{1/5}Ni_{1/5}Cu_{1/5}O$ | 2.246 ± **0.173** | **51.35** ± 0.98 | ✖ | -- |
| $Mg_{1/5}Ca_{1/5}Fe_{1/5}Co_{1/5}Cu_{1/5}O$ | 2.232 ± **0.174** | **67.15** ± 1.03 | ✖ | -- |
| $Mg_{1/5}Ca_{1/5}Mn_{1/5}Co_{1/5}Cu_{1/5}O$ | 2.241 ± **0.177** | **63.91** ± 0.93 | ✖ | -- |
| $Ca_{1/6}Mn_{1/6}Fe_{1/6}Co_{1/6}Cu_{1/6}Zn_{1/6}O$ | 2.246 ± **0.178** | **66.13** ± 0.88 | ✖ | -- |
| $Mg_{1/5}Ca_{1/5}Mn_{1/5}Fe_{1/5}Cu_{1/5}O$ | 2.252 ± **0.178** | **52.61** ± 1.08 | ✖ | -- |
| $Ca_{1/5}Co_{1/5}Ni_{1/5}Cu_{1/5}Zn_{1/5}O$ | 2.224 ± **0.179** | **99.96** ± 0.77 | ✖ | -- |
| $Mg_{1/5}Ca_{1/5}Co_{1/5}Cu_{1/5}Zn_{1/5}O$ | 2.229 ± **0.180** | **94.66** ± 0.91 | ✖ | -- |
| $Ca_{1/5}Mn_{1/5}Fe_{1/5}Co_{1/5}Cu_{1/5}O$ | 2.254 ± **0.182** | **47.07** ± 0.89 | ✖ | -- |
| $Ca_{1/5}Fe_{1/5}Ni_{1/5}Cu_{1/5}Zn_{1/5}O$ | 2.237 ± **0.188** | **82.80** ± 1.04 | ✖ | -- |
| $Ca_{1/4}Co_{1/4}Ni_{1/4}Cu_{1/4}O$ | 2.230 ± **0.188** | **86.78** ± 0.80 | ✖ | -- |
| $Mg_{1/5}Ca_{1/5}Fe_{1/5}Cu_{1/5}Zn_{1/5}O$ | 2.242 ± **0.193** | **74.54** ± 1.23 | ✖ | -- |
| $Ca_{1/5}Mn_{1/5}Ni_{1/5}Cu_{1/5}Zn_{1/5}O$ | 2.248 ± **0.194** | **77.54** ± 0.98 | ✖ | -- |
| $Ca_{1/4}Fe_{1/4}Ni_{1/4}Cu_{1/4}O$ | 2.245 ± **0.195** | **67.12** ± 1.24 | ✖ | -- |
| $Mg_{1/4}Ca_{1/4}Co_{1/4}Cu_{1/4}O$ | 2.241 ± **0.196** | **86.53** ± 1.04 | ✖ | -- |
| $Mg_{1/5}Ca_{1/5}Mn_{1/5}Cu_{1/5}Zn_{1/5}O$ | 2.252 ± **0.198** | **75.93** ± 1.01 | ✖ | -- |
| $Ca_{1/4}Mn_{1/4}Ni_{1/4}Cu_{1/4}O$ | 2.258 ± **0.200** | **57.55** ± 0.82 | ✖ | -- |
| $Mg_{1/4}Ca_{1/4}Fe_{1/4}Cu_{1/4}O$ | 2.255 ± **0.203** | **61.73** ± 1.50 | ✖ | -- |
| $Mg_{1/4}Ca_{1/4}Mn_{1/4}Cu_{1/4}O$ | 2.266 ± **0.205** | **61.51** ± 1.18 | ✖ | -- |
| $Ca_{1/5}Fe_{1/5}Co_{1/5}Cu_{1/5}Zn_{1/5}O$ | 2.249 ± **0.205** | **76.86** ± 0.92 | ✖ | -- |
| $Mg_{1/4}Ca_{1/4}Cu_{1/4}Zn_{1/4}O$ | 2.252 ± **0.211** | **104.69** ± 0.97 | ✖ | -- |
| $Ca_{1/4}Ni_{1/4}Cu_{1/4}Zn_{1/4}O$ | 2.247 ± **0.212** | **113.67** ± 0.95 | ✖ | -- |
| $Ca_{1/5}Mn_{1/5}Co_{1/5}Cu_{1/5}Zn_{1/5}O$ | 2.260 ± **0.212** | **72.61** ± 1.05 | ✖ | -- |
| $Ca_{1/4}Fe_{1/4}Co_{1/4}Cu_{1/4}O$ | 2.260 ± **0.214** | **57.71** ± 1.08 | ✖ | -- |
| $C_{1/5}Mn_{1/5}Fe_{1/5}Cu_{1/5}Zn_{1/5}O$ | 2.272 ± **0.215** | **62.27** ± 1.02 | ✖ | -- |
| $Ca_{1/4}Mn_{1/4}Fe_{1/4}Cu_{1/4}O$ | 2.287 ± **0.219** | **37.87** ± 1 .09 | ✖ | -- |
| $Ca_{1/4}Mn_{1/4}Co_{1/4}Cu_{1/4}O$ | 2.273 ± **0.221** | **49.15** ± 0.96 | ✖ | -- |
| $Ca_{1/4}Co_{1/4}Cu_{1/4}Zn_{1/4}O$ | 2.268 ± **0.246** | **96.59** ± 0.83 | ✖ | -- |
| $Ca_{1/4}Fe_{1/4}Cu_{1/4}Zn_{1/4}O$ | 2.285 ± **0.259** | **73.25** ± 1.35 | ✖ | -- |
| $Ca_{1/4}Mn_{1/4}Cu_{1/4}Zn_{1/4}O$ | 2.299 ± **0.265** | **71.34** ± 1.02 | ✖ | -- |

## Methods

*Computational*

CHGNet version 0.3.5 was used for all calculations. SQSs for benchmarking were constructed using the integrated cluster expansion toolkit [18,28]. All atomic coordinates were initially randomly rattled with the atomic simulation environment [29] to allow symmetry-breaking during structural relaxation procedure [30]. We find that this is important to realize the Jahn-Teller distortions for both CHGNet and DFT calculations. Structural relaxation was performed until forces were less than 50 meV/Å. We note that only the ground-state (0 K) structural predictions and corresponding energies are explored here, while CHGNet is also capable of performing finite temperature simulations. We utilize the Warren-Cowley short-range ordering parameter [31] to quantify the randomness of our sampled structures. Pymatgen was utilized for analysis [22].

The Vienna Ab-initio Software Package (VASP) 6.4.1 was used for DFT calculations with the projector augmented wave pseudopotentials [32]. Calculation parameters were largely unchanged from Materials Project *MPRelaxSet* [22,33] defaults and magnetic ordering was initialized as ferromagnetic to ensure compatibility with the calculations on which CHGNet was fitted. The mixing scheme for calculations including a Hubbard U was also used [23]. Relaxed structures from CHGNet structure optimization were used as inputs for the DFT calculations to expedite the benchmarking procedure. For a consistent comparison with CHGNet relaxations, forces were minimized to less than 50 meV/Å.

*Experimental*

X-ray diffraction (XRD) was utilized before weighing the powders to ensure that CaO had not reacted with C to form $CaCO_3$, hence avoiding the need for a calcination step. XRD was done on PANalytical Empyrean diffractometer in Bragg-Brentano High-Definition Mode. Then we weighed Millipore-Sigma CaO, MgO, NiO, CuO, CoO, and ZnO to accurately give the intended composition with different Ca concentrations. The powder mix is then blended in a speed mixer followed by a shaker mill along with 3-mm and 5-mm diameter yttrium-stabilized zirconia milling media. To ensure thorough mixing, each batch was milled for a minimum of 2.5 hours. The mixed powders were then divided into 0.500-g pellets and compressed into diameter of 1.27 cm using a uniaxial hydraulic press exerting 20,000 psi. We subsequently scanned the unreacted pellets with both XRD and calibrated PANalytical Epsilon 1 X-ray fluorescence (XRF) to monitor and compare phases and composition before and after sintering. Phases were identified using PANalytical HighScore with 2024 updated materials library. The pellets were fired in air at 875, 950, 1000 and 1100°C for 24 hours. Pellets reacted at 950°C (Figure 4) are most representative of the system stability as 950°C is the highest sintering temperature accessible in air before CuO melting onsets. XRF also confirmed that we maintained the desired composition with 1-2% deviations after sintering at 950°C.

## References

[1] B. Cantor, I.T.H. Chang, P. Knight, A.J.B. Vincent, Microstructural development in equiatomic multicomponent alloys, Materials Science and Engineering: A 375–377 (2004) 213–218. https://doi.org/10.1016/j.msea.2003.10.257.

[2] J.-W. Yeh, S.-K. Chen, S.-J. Lin, J.-Y. Gan, T.-S. Chin, T.-T. Shun, C.-H. Tsau, S.-Y. Chang, Nanostructured High-Entropy Alloys with Multiple Principal Elements: Novel Alloy Design Concepts and Outcomes, Advanced Engineering Materials 6 (2004) 299–303. https://doi.org/10.1002/adem.200300567.

[3] C.M. Rost, E. Sachet, T. Borman, A. Moballegh, E.C. Dickey, D. Hou, J.L. Jones, S. Curtarolo, J.-P. Maria, Entropy-stabilized oxides, Nat Commun 6 (2015) 8485. https://doi.org/10.1038/ncomms9485.

[4] A. Sarkar, L. Velasco, D. Wang, Q. Wang, G. Talasila, L. de Biasi, C. Kübel, T. Brezesinski, S.S. Bhattacharya, H. Hahn, B. Breitung, High entropy oxides for reversible energy storage, Nat Commun 9 (2018) 3400. https://doi.org/10.1038/s41467-018-05774-5.

[5] Y. Son, W. Zhu, S.E. Trolier-McKinstry, Electrocaloric Effect of Perovskite High Entropy Oxide Films, Advanced Electronic Materials 8 (2022) 2200352. https://doi.org/10.1002/aelm.202200352.

[6] M. Brahlek, A.R. Mazza, K.C. Pitike, E. Skoropata, J. Lapano, G. Eres, V.R. Cooper, T.Z. Ward, Unexpected crystalline homogeneity from the disordered bond network in La ( C r 0.2 M n 0.2 F e 0.2 C o 0.2 N i 0.2 ) O 3 films, Phys. Rev. Materials 4 (2020) 054407. https://doi.org/10.1103/PhysRevMaterials.4.054407.

[7] G.N. Kotsonis, S.S.I. Almishal, L. Miao, M.K. Caucci, G.R. Bejger, S.V.G. Ayyagari, T.W. Valentine, B.E. Yang, S.B. Sinnott, C.M. Rost, N. Alem, J.-P. Maria, Fluorite-structured high-entropy oxide sputtered thin films from bixbyite target, Applied Physics Letters 124 (2024) 171901. https://doi.org/10.1063/5.0201419.

[8] R. Djenadic, A. Sarkar, O. Clemens, C. Loho, M. Botros, V.S.K. Chakravadhanula, C. Kübel, S.S. Bhattacharya, A.S. Gandhi, H. Hahn, Multicomponent equiatomic rare earth oxides, Materials Research Letters 5 (2017) 102–109. https://doi.org/10.1080/21663831.2016.1220433.

[9] P. Sarker, T. Harrington, C. Toher, C. Oses, M. Samiee, J.-P. Maria, D.W. Brenner, K.S. Vecchio, S. Curtarolo, High-entropy high-hardness metal carbides discovered by entropy descriptors, Nat Commun 9 (2018) 4980. https://doi.org/10.1038/s41467-018-07160-7.

[10] S. Divilov, H. Eckert, D. Hicks, C. Oses, C. Toher, R. Friedrich, M. Esters, M.J. Mehl, A.C. Zettel, Y. Lederer, E. Zurek, J.-P. Maria, D.W. Brenner, X. Campilongo, S. Filipović, W.G. Fahrenholtz, C.J. Ryan, C.M. DeSalle, R.J. Crealese, D.E. Wolfe, A. Calzolari, S. Curtarolo, Disordered enthalpy–entropy descriptor for high-entropy ceramics discovery, Nature 625 (2024) 66–73. https://doi.org/10.1038/s41586-023-06786-y.

[11] K.C. Pitike, S. Kc, M. Eisenbach, C.A. Bridges, V.R. Cooper, Predicting the Phase Stability of Multicomponent High-Entropy Compounds, Chem. Mater. 32 (2020) 7507–7515. https://doi.org/10.1021/acs.chemmater.0c02702.

[12] S.S. Aamlid, G.H.J. Johnstone, S. Mugiraneza, M. Oudah, J. Rottler, A.M. Hallas, Phase stability of entropy stabilized oxides with the α-PbO2 structure, Commun Mater 4 (2023) 45. https://doi.org/10.1038/s43246-023-00372-5.

[13] Y. Pu, D. Moseley, Z. He, K.C. Pitike, M.E. Manley, J. Yan, V.R. Cooper, V. Mitchell, V.K. Peterson, B. Johannessen, R.P. Hermann, P. Cao, (Mg,Mn,Fe,Co,Ni)O: A rocksalt high-entropy oxide containing divalent Mn and Fe, Sci. Adv. 9 (2023) eadi8809. https://doi.org/10.1126/sciadv.adi8809.

[14] S.S.I. Almishal, J.T. Sivak, G.N. Kotsonis, Y. Tan, M. Furst, D. Srikanth, V.H. Crespi, V. Gopalan, J.T. Heron, L.-Q. Chen, C.M. Rost, S.B. Sinnott, J.-P. Maria, Untangling individual cation roles in rock salt high-entropy oxides, Acta Materialia (2024) 120289. https://doi.org/10.1016/j.actamat.2024.120289.

[15] C.J. Bartel, A.W. Weimer, S. Lany, C.B. Musgrave, A.M. Holder, The role of decomposition reactions in assessing first-principles predictions of solid stability, Npj Comput Mater 5 (2019) 4. https://doi.org/10.1038/s41524-018-0143-2.

[16] W. Hume-Rothery, G.W.M. Abbott, THE FREEZING POINTS. MELTING POINTS, AND SOLID SOLUBILITY LIMITS OF THE ALLOYS OF SILVER AND COPPER WITH THE ELEMENTS OF THE B SUB-GROUPS., (n.d.).

[17] A. Kretschmer, P.H. Mayrhofer, Explaining the entropy forming ability for carbides with the effective atomic size mismatch, Sci Rep 14 (2024) 7210. https://doi.org/10.1038/s41598-024-57456-6.

[18] A. Zunger, S.-H. Wei, L.G. Ferreira, J.E. Bernard, Special quasirandom structures, Phys. Rev. Lett. 65 (1990) 353–356. https://doi.org/10.1103/PhysRevLett.65.353.

[19] K. Yang, C. Oses, S. Curtarolo, Modeling Off-Stoichiometry Materials with a High-Throughput Ab-Initio Approach, Chem. Mater. 28 (2016) 6484–6492. https://doi.org/10.1021/acs.chemmater.6b01449.

[20] B. Deng, P. Zhong, K. Jun, J. Riebesell, K. Han, C.J. Bartel, G. Ceder, CHGNet as a pretrained universal neural network potential for charge-informed atomistic modelling, Nat Mach Intell 5 (2023) 1031–1041. https://doi.org/10.1038/s42256-023-00716-3.
[21] J. Riebesell, R.E.A. Goodall, A. Jain, P. Benner, K.A. Persson, A.A. Lee, Matbench Discovery -- An evaluation framework for machine learning crystal stability prediction, (2023). http://arxiv.org/abs/2308.14920 (accessed October 6, 2023).
[22] S.P. Ong, W.D. Richards, A. Jain, G. Hautier, M. Kocher, S. Cholia, D. Gunter, V.L. Chevrier, K.A. Persson, G. Ceder, Python Materials Genomics (pymatgen): A robust, open-source python library for materials analysis, Computational Materials Science 68 (2013) 314–319.
[23] A. Jain, G. Hautier, S.P. Ong, C.J. Moore, C.C. Fischer, K.A. Persson, G. Ceder, Formation enthalpies by mixing GGA and GGA + U calculations, Phys. Rev. B 84 (2011) 045115. https://doi.org/10.1103/PhysRevB.84.045115.
[24] D. Berardan, A.K. Meena, S. Franger, C. Herrero, N. Dragoe, Controlled Jahn-Teller distortion in (MgCoNiCuZn)O-based high entropy oxides, Journal of Alloys and Compounds 704 (2017) 693–700. https://doi.org/10.1016/j.jallcom.2017.02.070.
[25] Zs. Rák, J.-P. Maria, D.W. Brenner, Evidence for Jahn-Teller compression in the (Mg, Co, Ni, Cu, Zn)O entropy-stabilized oxide: A DFT study, Materials Letters 217 (2018) 300–303. https://doi.org/10.1016/j.matlet.2018.01.111.
[26] See Supplemental Material at [ ], (n.d.).
[27] R.D. Shannon, Revised effective ionic radii and systematic studies of interatomic distances in halides and chalcogenides, Foundations of Crystallography 32 (1976) 751–767.
[28] M. Ångqvist, W.A. Muñoz, J.M. Rahm, E. Fransson, C. Durniak, P. Rozyczko, T.H. Rod, P. Erhart, ICET – A Python Library for Constructing and Sampling Alloy Cluster Expansions, Adv. Theory Simul. 2 (2019) 1900015. https://doi.org/10.1002/adts.201900015.
[29] A. Hjorth Larsen, J. Jørgen Mortensen, J. Blomqvist, I.E. Castelli, R. Christensen, M. Dułak, J. Friis, M.N. Groves, B. Hammer, C. Hargus, E.D. Hermes, P.C. Jennings, P. Bjerre Jensen, J. Kermode, J.R. Kitchin, E. Leonhard Kolsbjerg, J. Kubal, K. Kaasbjerg, S. Lysgaard, J. Bergmann Maronsson, T. Maxson, T. Olsen, L. Pastewka, A. Peterson, C. Rostgaard, J. Schiøtz, O. Schütt, M. Strange, K.S. Thygesen, T. Vegge, L. Vilhelmsen, M. Walter, Z. Zeng, K.W. Jacobsen, The atomic simulation environment—a Python library for working with atoms, J. Phys.: Condens. Matter 29 (2017) 273002. https://doi.org/10.1088/1361-648X/aa680e.
[30] Y. Zhang, J. Furnes, R. Zhang, Z. Wang, A. Zunger, J. Sun, Symmetry-Breaking Polymorphous Descriptions for Complex Materials without Interelectronic U, Phys. Rev. B 102 (2020) 045112. https://doi.org/10.1103/PhysRevB.102.045112.
[31] J.M. Cowley, An Approximate Theory of Order in Alloys, Phys. Rev. 77 (1950) 669–675. https://doi.org/10.1103/PhysRev.77.669.
[32] G. Kresse, J. Furthmüller, Efficient iterative schemes for *ab initio* total-energy calculations using a plane-wave basis set, Phys. Rev. B 54 (1996) 11169–11186. https://doi.org/10.1103/PhysRevB.54.11169.
[33] A. Jain, S.P. Ong, G. Hautier, W. Chen, W.D. Richards, S. Dacek, S. Cholia, D. Gunter, D. Skinner, G. Ceder, K.A. Persson, Commentary: The Materials Project: A materials genome approach to accelerating materials innovation, APL Materials 1 (2013) 011002. https://doi.org/10.1063/1.4812323.